\documentclass[11pt,a4paper]{article}
\usepackage{jheppub}

\usepackage{amsmath}
\usepackage{amsfonts}
\usepackage{amsthm}
\usepackage{slashed}
\usepackage{graphicx}
\usepackage{mathrsfs,bbm}
\usepackage{subfig}
\usepackage{booktabs}
\usepackage[numbers,sort&compress]{natbib}
\usepackage[utf8]{inputenc}

\usepackage{multirow}

\makeatletter
\g@addto@macro\bfseries{\boldmath}
\makeatother

\theoremstyle{remark}

\newcommand{\eps} {\epsilon}

\newcommand{\OLL}{{\rmfamily\scshape OpenLoops2}}

\newcommand{\fin}{{\rm fin}}

\title{Two-loop QCD corrections to the 
$V\to q\bar{q}g$ helicity amplitudes with axial-vector couplings}

\author[a]{Thomas Gehrmann} 
\author[b]{, Tiziano Peraro} 
\author[c]{and Lorenzo Tancredi}

\affiliation[a]{Physik-Institut, Universität Zürich, Winterthurerstrasse 190, CH-8057 Zürich, Switzerland} 
\affiliation[b]{Dipartimento di Fisica e Astronomia, Università di Bologna e INFN, Sezione di Bologna, via
Irnerio 46, I-40126 Bologna, Italy}
\affiliation[c]{Physik Department, James-Franck-Straße 1, Technische Universität München,
D–85748 Garching, Germany}

\emailAdd{thomas.gehrmann@uzh.ch}
\emailAdd{tiziano.peraro@unibo.it}
\emailAdd{lorenzo.tancredi@tum.de}

\abstract{
We compute the two-loop corrections to the 
helicity amplitudes for the coupling of a 
massive vector boson to a massless quark-antiquark pair and a gluon, accounting for vector and axial-vector couplings of the vector boson
and distinguishing isospin non-singlet and 
singlet contributions. 
A new four-dimensional basis for the 
decomposition of the amplitudes into 12 invariant tensor structures is introduced. The 
associated form factors are then computed 
up to two loops in QCD using 
dimensional regularization. 
After performing renormalization and infrared subtraction, the finite parts of the renormalized
non-singlet vector and axial-vector form factors are shown agree with each other, and to reproduce the previously known two-loop amplitudes. The singlet axial-vector amplitude receives a contribution from the axial anomaly from two loops onwards. This amplitude is computed for massless and massive internal quarks. Our results provide the last missing 
two-loop amplitudes entering the NNLO QCD corrections of vector-boson-plus-jet production at hadron colliders. 
}

\preprint{\begin{minipage}[t]{8cm}\begin{flushright} 
TUM-HEP-1439/22\\
ZU-TH 49/22\\
      \end{flushright}\end{minipage}}

\begin{document}

\maketitle

\catcode`\@=11
\font\manfnt=manfnt
\def\Watchout{\@ifnextchar [{\W@tchout}{\W@tchout[1]}}
\def\W@tchout[#1]{{\manfnt\@tempcnta#1\relax%
  \@whilenum\@tempcnta>\z@\do{%
    \char"7F\hskip 0.3em\advance\@tempcnta\m@ne}}}
\let\foo\W@tchout
\def\dubious{\@ifnextchar[{\@dubious}{\@dubious[1]}}
\let\enddubious\endlist
\def\@dubious[#1]{%
  \setbox\@tempboxa\hbox{\@W@tchout#1}
  \@tempdima\wd\@tempboxa
  \list{}{\leftmargin\@tempdima}\item[\hbox to 0pt{\hss\@W@tchout#1}]}
\def\@W@tchout#1{\W@tchout[#1]}
\catcode`\@=12


\section{Introduction}

The scattering amplitudes involving a massive electroweak gauge boson and three massless QCD partons enter, in their different kinematical crossings, the theory predictions for important QCD precision observables: $e^+e^- \to 3$~jets, $ep\to (2+1)$~jets and $pp\to V+$jet.  

Especially 
three-jet prodction in $e^+e^-$ played an outstanding role in establishing QCD as theory of the strong interaction and in the discovery of the gluon~\cite{Barber:1979yr}. The quest for a precise theoretical understading of three-jet production observables~\cite{Ellis:1976uc} 
demanded higher order 
perturbative QCD corrections. The one-loop corrections to the $V\to q\bar qg$ amplitudes were computed in the context of the 
next-to-leading order (NLO) corrections to three-jet production~\cite{Ellis:1980wv,Giele:1991vf}, and the two-loop corrections to these 
amplitudes~\cite{Garland:2001tf,Garland:2002ak} were 
first applied in computing the next-to-next-to-leading order (NNLO) corrections to this 
process~\cite{Gehrmann-DeRidder:2007vsv,Gehrmann-DeRidder:2008qsl,Weinzierl:2009ms,DelDuca:2016csb}. Crossings of the scattering amplitudes to 
the other processes were then obtained by analytic continuation~\cite{Graudenz:1993tg,Gehrmann:2002zr} to the relevant kinematical regions.

\begin{figure}[t!]
    \centering
    \includegraphics[scale=0.5]{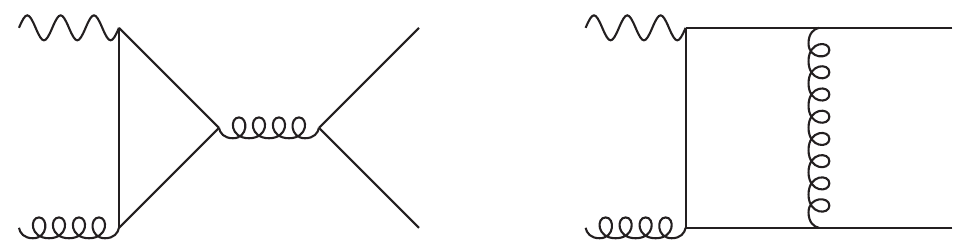}
    \caption{Example of Feynman diagrams for singlet (left) and non-singlet (right) contributions.}
    \label{fig:fig1}
\end{figure}

Beyond tree-level, one can separate the $V\to q\bar qg$ amplitudes by an isospin projection on the external quarks into non-singlet contributions (where the gauge boson couples to the spin line of the external quark) and pure-singlet contributions (where the gauge boson couples to a closed quark loop unrelated to the external quark spin line).  Example diagrams are shown in Figure~\ref{fig:fig1}. The pure-singlet contributions vanish trivially for $V=W^\pm$ due to charge conservation.

The coupling structure of the external 
vector boson to 
quarks involves a vector and an axial-vector component. For massless external quarks, one expects the  non-singlet contributions to the vector and axial-vector $V\to q\bar qg$
amplitudes to agree due to chirality conservation of massless fermions.
Owing to the conceptual difficulty of handling axial-vector currents in dimensional 
regualrization, this agreement has however never been checked explicitly beyond tree level.

The pure-singlet contribution at one loop vanishes for a vector coupling. 
For an axial-vector 
coupling, one finds a finite one-loop amplitude~\cite{Dicus:1985wx,Hagiwara:1990dx}, which vanishes if the internal quark flavours are summed over a mass-degenerate isospin doublet. Consequently, the large mass splitting between top and bottom quarks in the third generation results in a non-vanising pure-singlet axial-vector contribution. The one-loop pure-singlet 
axial-vector amplitude is finite for internal massless quarks $q'$, and suppressed as $(m_{Q'}/m_V)^2$  in the 
large mass limit of the internal quark $Q'$. 
At two loops, the pure-singlet contribution 
was computed up to now only for a vector coupling~\cite{Garland:2001tf,Garland:2002ak}.

Most jet observables in $e^+e^-$ average over 
the incoming beam direction. In this case, the 
pure-singlet axial-vector contribution averages out to zero. Its contribution to jet production in $ep$ collisions is strongly suppressed by the $Z$ boson mass. In $Z+$jet production, the 
numerical magnitude of the one-loop pure-singlet axial-vector contribution was found to be very small (per-mille correction) for 
sufficiently inclusive observables~\cite{Dicus:1985wx}, but potentially becomes sizable in specific angular correlations between the jet and the decay lepton momenta~\cite{Hagiwara:1991xy}. 

Up to now, all NNLO QCD calculations of $Z+$jet production at the LHC~\cite{Gehrmann-DeRidder:2015wbt,Boughezal:2015ded,Neumann:2022lft} discarded the 
axial-vector pure-singlet contributions, due to the lack of the corresponding two-loop axial-vector amplitudes for crossings of $Z\to q\bar q g$. At the same level, other axial-vector pure-singlet contributions to $Z+$jet production
are known~\cite{vanderBij:1988ac,Hopker:1993pb,Bern:1997sc}, but could not be 
included consistently up to now. 
The very same axial-vector pure-singlet amplitudes also contribute to the third-order (N3LO) corrections to $Z$ boson production at hadron colliders. Their 
contribution to the inclusive $Z$ boson coefficient functions at N3LO has been computed
most recently~\cite{Duhr:2021vwj} by combining 
loop and phase space integrations. 

It is the purpose of the current paper to complete the two-loop helicity amplitudes
for $Z\to q\bar q g$ by explicitly establishing the identity of vector and axial-vector 
non-singlet amplitudes and by computing the missing pure-singlet axial-vector amplitudes for massless and very massive internal quarks. 
This paper is structured as follows: in Section~\ref{sec:tensor}, we describe a novel four-dimensional tensor decomposition of the helicity amplitudes and introduce projectors on the relevant form factors, which are then computed in Section~\ref{sec:ff}. Their renormalization and infrared factorization is described in Section~\ref{sec:renorm}. The
form factors are then assembled into the helicity amplitudes for $Z\to q\bar q g$  in Section~\ref{sec:helamp}. 
 We conclude with an outlook in Section~\ref{sec:conc}. 


\section{The tensor decomposition including axial-vector terms}
\label{sec:tensor}
We consider the decay of a space-like 
electroweak gauge boson to a pair of quarks and
a gluon up to two loops in QCD.
We work in massless QCD assuming 
$N_f=5$ massless quarks. 
When evaluated for a single quark flavour, the axial-vector pure-singlet contribution contains an 
axial anomaly, which is cancelled upon summation over both quark flavours in 
 an isospin doublet. 
 To obtain a consistent result, we also allow 
 top quarks in the virtual loops of the 
axial-vector pure-singlet contribution (and only in this contribution), 
which we compute in the limit of   infinitely large top-mass $m_t \to \infty$. 
As a byproduct of our calculation, 
we also recompute the well-known
non-singlet contributions, limiting ourselves in that case to consider only massless quarks
circulating in the loops.

The full electroweak Standard Model couplings are inserted only in Section~\ref{sec:helamp} when assembling the 
helicity amplitudes. For their tensor decomposition and the subsequent calculation of the 
corresponding form factors, we instead consider QCD coupled to an external vector or
 axial-vector current with unit coupling constant. The vector current is conserved to all orders 
in perturbation theory, and in any number of dimensions. The pure-singlet axial-vector current is anomalous and requires 
renormalization (which cancels out upon summation over a quark isospin doublet). 
The conservation of the non-singlet axial-vector current is broken by dimensional 
regularisation and has to be restored by a finite renormalization. The renormalization of 
axial-vector currents in QCD is described in detail in Section~\ref{sec:renorm} below.

For definiteness, we work in the decay kinematics
$$V(p_4) \to \bar{q}(p_1) + q(p_2) + g(p_3)\,,$$
from which the relevant scattering channels can be obtained by
crossing symmetry and a corresponding analytic continuation~\cite{Gehrmann:2002zr}.
To parametrise the kinematics of this scattering process, 
we recall that 
$$p_1^2 =p_2^2=p_3^2 = 0\,, \quad p_4^2 = (p_1+p_2+p_3)^2 = q^2\,,$$
and  introduce the  usual Mandelstam invariants, defined in the decay
kinematics as
\begin{equation}
    s_{12}=(p_1+p_2)^2\,, \quad s_{13}=(p_1+p_3)^2\,, \quad 
    s_{23}=(p_2+p_3)^2\,, \quad s_{12}+s_{13}+s_{23} = q^2\,.
\end{equation}
Note that for generality we use $q^2$ for the invariant mass of the vector boson.
For later convenience, following~\cite{Gehrmann:2000zt,Gehrmann:2001ck} we also introduce the three dimensionless ratios
\begin{equation}
    x = \frac{s_{12}}{q^2}\,, \quad y = \frac{s_{13}}{q^2}\,, \quad z = \frac{s_{23}}{q^2}\,, \label{eq:ratios}
    \quad \mbox{with} \quad x+y+z=1\,.
\end{equation}
In terms of any pair of the variables above, say $y,z$, the decay kinematics is mapped by $$0<y<1 \,, \quad 0<z<1-y\,.$$

A possible approach to the computation of multi-loop amplitudes 
starts with their decomposition into a basis of independent
Lorentz tensor structures, from which one can then extract
the so-called helicity amplitudes. 
Our starting point here is the construction described in~\cite{Peraro:2019cjj,Peraro:2020sfm}, which makes it possible to
avoid the calculation of evanescent tensor structures in $d=4$ space-time dimensions. As we will see, this is particularly
convenient in the presence of an axial coupling.
A nice aspect of this construction is also that
the number of independent tensor structures always matches one-to-one
the number of independent helicity amplitudes in $d=4$. 

In the process under consideration
helicity conservation along the massless external fermion line implies that
there can be $2\times2\times3=12$ independent helicity configurations:
$2$ for the quark line, $2$ for the on-shell gluon and $3$ for the
off-shell vector boson $V$. In the absence of parity breaking terms, 
these would be reduced to $6$ by the trivial transformation properties
of the helicity amplitudes of a $2 \to 2$ scattering process under 
parity. As explained in~\cite{Peraro:2019cjj}, this reduction no longer takes place
starting for five or more external particles, since the helicity amplitudes in that case depend explicitly 
on the parity-breaking object 
${\rm tr}_5 = \epsilon^{p_1 p_2 p_3 p_4}$. 
Clearly,
if the vector boson couples chirally as a Standard Model $Z$ or $W$ boson, 
this simplification does not occur and we
expect a total of $12$ independent tensor structures to be required to
fully decompose both vector and axial-vector parts of the scattering amplitude.

As it was shown in~\cite{Peraro:2019cjj,Peraro:2020sfm}, 
by working in the 't Hooft-Veltman 
dimensional regularisation scheme~\cite{'tHooft:1972fi},
we can limit ourselves to perform a tensor decompositon
assuming that all external states are in  exactly 
$d=4$ space-time
dimensions,
since all remaining evanescent tensor structures turn out
not to contribute to the helicity amplitudes, even in $d=4-2 \epsilon$.
To construct the tensor decomposition for our problem, 
a convenient starting point is
a basis of four independent vectors in $d=4$. 
The first three vectors can
be chosen to be the three independent momenta $p_1^\mu$, $p_2^\mu$,
$p_3^\mu$, such that the natural fourth independent vector 
is the orthogonal, axial-vector
\begin{equation}
\eps^{\nu \rho \sigma \mu} p_{1\nu} p_{2\rho} p_{3\sigma} = \eps^{p_1 p_2 p_3 \mu} = v_A^\mu, 
\label{eq:defva}
\end{equation}
where $\eps^{\mu \nu \rho \sigma}$
is the Levi-Civita symbol, defined using \textsc{Form} conventions that correspond to
\begin{equation}
    \epsilon^{0123}=- \epsilon_{0123}=-i, \qquad \epsilon^{\mu \nu \rho \sigma}     \epsilon_{\mu \nu \rho \sigma} = 24 + \mathcal{O}(d-4).
\end{equation}
We stress here that $p_i \cdot v_A = 0$.

Note also that, by using $q_i^\mu = \{ p_1^\mu, p_2^\mu, p_3^\mu, v_A^\mu \}$ 
as independent vectors, we do not need to include in our
tensor decomposition
 neither $g^{\mu \nu}$ nor other higher-rank tensors obtained from the 
 Levi-Civita tensor as $\eps^{p_i p_j \mu \nu}$ for $i,j =1,2,3$, 
since in $d=4$ they are not linearly independent from
all tensors built from the $q_i^\mu$.
A crucial point here, since we are dealing with external fermions, 
is that the same is true also for 
the Dirac $\gamma^\mu$, and we have
\begin{align}
\gamma^\mu = \sum_{i=1}^4 \hat{a}_i q_i^\mu\,, \quad 
g^{\mu \nu} = \sum_{i,j=1}^{4} b_{ij} q_i^\mu q_j^\mu\,, \quad
\eps^{p_k p_l \mu \nu} = \sum_{i,j=1}^{4} c_{ij}^{(kl)} q_i^\mu q_j^\mu\,,
\end{align}
where clearly the coefficients $b_{ij}$ and $c_{ij}^{(kl)}$ 
are scalar functions while
$\hat{a}_i$ must be linear combinations of $\slashed{q}_i = q_i^{\mu} \gamma_\mu$. 
Their exact form does not matter for what follows.

With these observations, and assuming that helicity
is conserved along the fermion line, the amplitude
can be easily decomposed in independent tensors.
We start by defining
\begin{equation}
    \mathcal{M} = -i\,\sqrt{4 \pi \bar{\alpha}_s} \; \mathbb{T}^a_{ij}\, \eps_{4,\mu} \eps_{3,\nu} A^{\mu\nu}
    \label{eq:ampdef}
\end{equation}
where $\bar{\alpha}_s$ is the bare strong coupling and $\mathbb{T}^a_{ij}$ are the ${\rm SU}(3)$ color fundamental
generators. 
The external vector or axial-vector vertex does not carry a coupling constant. 
The rank-two tensor can be decomposed as
\begin{align}
 A^{\mu\nu} &= 
\bar{u}(p_2) \slashed{p}_3 u(p_1) 
\Big[  \tilde{F}_1 p_1^\mu p_1^\nu  + \tilde{F}_2 p_2^\mu p_1^\nu  + \tilde{F}_3 v_A^\mu v_A^\nu
 + \tilde{G}_1 p_1^\mu v_A^\nu + \tilde{G}_2 p_2^\mu  v_A^\nu + \tilde{G}_3 v_A^\mu p_1^\nu
\Big]
\nonumber \\
&
+\bar{u}(p_2) \slashed{v}_A u(p_1) 
\Big[  
\tilde{F}_4 p_1^\mu v_A^\nu + \tilde{F}_5 p_2^\mu  v_A^\nu + \tilde{F}_6 v_A^\mu p_1^\nu
+\tilde{G}_4 p_1^\mu p_1^\nu  + \tilde{G}_5 p_2^\mu p_1^\nu  + \tilde{G}_6 v_A^\mu v_A^\nu
\Big], \label{eq:tensdecA}
\end{align}
where $\tilde{F}_i$ and $\tilde{G}_i$ are scalar form factors.
In order to obtain~\eqref{eq:tensdecA}, we also used the transversality
condition for the external gluon
$$\eps_3 \cdot p_3 = 0\,,$$
together with the following gauce choices for the gluon and the off-shell
vector boson
$$\eps_3 \cdot p_2 = 0\,, \qquad \epsilon_4 \cdot p_4 = 0 \quad \textrm{ (Lorenz Gauge) } .$$
We stress that this gauge choice implies 
the following polarisation sums rules
\begin{align}
\sum_{pol} \epsilon_3^{\mu*} \epsilon_3^\nu = 
- g^{\mu \nu} + \frac{p_3^\mu p_{2}^\nu + p_3^\nu p_{2}^\mu }{p_2 \cdot p_3}\,,
\qquad
\sum_{pol} \epsilon_4^{\mu*} \epsilon_4^\nu = 
- g^{\mu \nu} + \frac{p_4^\mu  p_4^\nu  }{q^2}\,.
\label{eq:polsumqqgg}
\end{align}
While fixing the gauge is not necessary in principle, 
it provides a clear way to enumerate all
independent structures.

Looking again at~\eqref{eq:tensdecA}, 
we see that by using the vector $v_A^\mu$, we generate
exactly $12$ independent tensors in $d=4$ dimensions, 
which conveniently separate into
$6$ parity-even ($\tilde{F}_i$) and $6$ parity-odd ($\tilde{G}_i$) structures.
We stress once more that this simple separation is only possible
for the scattering of up to $4$ particles.

Starting from~\eqref{eq:tensdecA},
we can obtain an alternative decomposition, 
that has the advantage of involving at most
one occurrence of the parity-odd vector $v_A^\mu$, only in those 
tensors that are parity-violating. 
 This is particularly useful to get rid of the possible 
ambiguity in the order of contraction
of pairs of Levi-Civita tensors when we define the corresponding projectors
and apply them on the amplitude using Larin scheme~\cite{Larin:1993tq}.
Starting from the fact that  $g^{\mu\nu}$
is parity-even (and can thus contains only products of an even number of $v_A^\mu$), we can write
$$g^{\mu\nu} = \sum_{i,j=1}^3 b_{ij} p_i^\mu p_j^\nu 
+ b\, v_A^\mu v_A^\nu\,,$$ 
and we easily see that we can effectively substitute 
$v_A^\mu v_A^\nu \sim g^{\mu \nu}$ everywhere in the tensor
decomposition, still spanning the same vector space. This provides the 
alternative tensor decomposition
\begin{align}
A^{\mu\nu} &= 
\bar{u}(p_2) \slashed{p}_3 u(p_1) 
\Big[  F_1 p_1^\mu p_1^\nu  + F_2 p_2^\mu p_1^\nu  + F_3 g^{\mu \nu}
 + G_1 p_1^\mu v_A^\nu + G_2 p_2^\mu  v_A^\nu + G_3 v_A^\mu p_1^\nu
\Big]
\nonumber \\
& + \bar{u}(p_2) \gamma^\nu u(p_1) 
\Big[  F_4 p_1^\mu  + F_5 p_2^\mu  \Big]  + \bar{u}(p_2) \gamma^\mu u(p_1)  F_6 p_1^\nu
\nonumber \\
&  
+\bar{u}(p_2) \slashed{v}_A u(p_1) 
\Big[   G_4 p_1^\mu p_1^\nu  + G_5 p_2^\mu p_1^\nu \Big]
+ G_6 \Big[ \bar{u}(p_2) \gamma^\mu u(p_1)  v_A^\nu +
\bar{u}(p_2) \gamma^\nu u(p_1)  v_A^\mu \Big]
, \label{eq:AAV}
\end{align}
in terms of new form factors $F_i$ and $G_j$.
The choice made for the tensor multiplying form factor $G_6$ deserves special attention. 
Indeed,
there would be three equivalent choices to substitute $v_A^\mu v_A^\nu$
$$\bar{u}(p_2) \slashed{v}_A u(p_1)   v_A^\mu v_A^\nu \to 
\left\{ \begin{array}{c}
\bar{u}(p_2) \slashed{v}_A u(p_1)   g^{\mu \nu} \\
\bar{u}(p_2) \gamma^\mu u(p_1)  v_A^\nu \\
\bar{u}(p_2) \gamma^\nu u(p_1)  v_A^\mu
\end{array} \right.\,,
$$
and the last two can be rearranged into their 
symmetric combination
\begin{equation} \bar{u}(p_2) \slashed{v}_A u(p_1)   v_A^\mu v_A^\nu \to 
\bar{u}(p_2) \gamma^\mu u(p_1)  v_A^\nu +
\bar{u}(p_2) \gamma^\nu u(p_1)  v_A^\mu\,. \label{eq:symtens}
\end{equation}
We chose this last option in~\eqref{eq:AAV} as the most natural one
in the 't Hooft-Veltman scheme, as we will elaborate upon in Section~\ref{sec:Funren} below.

We rewrite~\eqref{eq:AAV} formally as
\begin{align}
A^{\mu\nu} &= 
\sum_{i=1}^6 F_i \,T_i^{E,\mu \nu} +
\sum_{i=1}^{6} G_i \,T_i^{O,\mu \nu}
, \label{eq:AAVformal}
\end{align}
 where the identification of the parity-even ($T_j^{E,\mu \nu}$) and 
 parity-odd ($T_j^{O,\mu \nu}$) tensors is obvious comparing with~\eqref{eq:AAV}.

The newly defined $12$ tensors should 
be thought of as vectors in a vector
space endowed with the scalar product defined by
\begin{align}
    T_i^{P_1\dagger} \cdot T_j^{P_2} = 
    T_i^{{P_1,{\mu_1} {\nu_1}}\dagger} \kappa_{\mu_1 \mu_2 \nu_1 \nu_2}
    T_j^{P_2,{\mu_2} {\nu_2}} \label{eq:scalprod}
\end{align}
where $P_i = \{E,O\}$ and the metric reads
\begin{equation}
    \kappa^{\mu_1 \mu_2 \nu_1 \nu_2} =
  \left( - g^{{\mu_1} {\mu_2}} + \frac{p_4^{\mu_1}  p_4^{\mu_2}  }{q^2} \right)
    \left( - g^{{\nu_1} {\nu_2}} 
    + \frac{p_3^{\nu_1} p_{2}^{\nu_2} + p_3^{\nu_2} p_{2}^{\nu_2} }{p_2 \cdot p_3}\right)\,.
\end{equation}
This definition implements the gauge choice imposed on the external vector bosons.
With this scalar product it is easy to see that
 even and odd tensors are mutually orthogonal
\begin{align}
    T_i^{E\dagger} \cdot T_j^{O} = T_i^{O\dagger} \cdot T_j^{E} = 0\,.
\end{align}
An important point should be stressed here: the decomposition
in ~\eqref{eq:AAV} should  not be interpreted as a purely 
four-dimensional
decomposition, but instead as a decomposition valid in the
't Hooft-Veltman scheme. This means that when we perform the
sums in~\eqref{eq:scalprod}, indices that are not explicitly
contracted with external momenta are taken to be $d$-dimensional.
This implies, in particular, that we use consistently throughout the calculation
\begin{equation}
g^{\mu \nu} g_{\mu \nu} = d\,, \quad v_A \cdot v_A = 
\frac{d-3}{4} s_{12} s_{13} s_{23}\,. \label{eq:ddimcontr}
\end{equation}
We stress here, to avoid a possible source of confusion with the Levi-Civita tensor, that when we
take the adjoint of the tensors, we \emph{do not} 
complex conjugate the vector $v_A$.

Thanks to the orthogonality of the two sets of tensors, 
we can define two independent sets of \emph{projector operators} $\mathcal{P}_i^{P}$,
which are suitable vectors in the dual vector space 
defined such that
\begin{align}
   \mathcal{P}_i^{P} \cdot T_j^{P} = \delta_{ij} \quad \mbox{for} \quad
   P=\{E,O\}\,.
\end{align}
By expanding the projectors in the same basis of dual vectors
\begin{align}
    \mathcal{P}_i^{P} = \sum_{j=1}^6 c_j^{(i),P} T_j^{P\dagger}\,, \label{eq:proj}
\end{align}
their coefficients $c_j^{(i),P}$ can be obtained by inverting the matrix
$(M^{P})_{nm} = T_n^{P\dagger} \cdot T_m^{P} $ as
\begin{equation}
    c_j^{(i),P} = \left( M^{P} \right)^{-1}_{ij}\,.
\end{equation}
We provide algebraic expressions for the  projectors as ancillary files.


\section{Calculation of the form factors}
\label{sec:ff}
Once the projector operators have been defined, it is conceptually
straightforward to apply them on a representation for the scattering
amplitude in terms of Feynman diagrams.
Our calculation proceeds here 
in a rather standard way: we produce all relevant diagrams at
tree level, one loop and two loops with QGRAF~\cite{Nogueira:1991ex}
and perform the necessary manipulations which follow from
the application of the projectors defined in~\eqref{eq:proj}
in FORM~\cite{Vermaseren:2000nd}.
To deal with $\gamma_5$ consistently with our tensor decomposition,
we use the Larin-scheme~\cite{Larin:1993tq}, which defines the axial-vector
current through the anti-symmetrised replacement
\begin{equation}
    \gamma^\mu \gamma^5 \to 
    \frac{1}{2}\left( \gamma^\mu \gamma^5 -  \gamma^5 \gamma^\mu\right)
    = \frac{1}{6} \epsilon^{\mu \nu \rho \sigma} \gamma_{\nu} \gamma_{\rho} \gamma_{\sigma}\,. \label{eq:larin}
\end{equation}
Notice that Larin prescription has been adapted to be consistent
with our conventions for the Levi-Civita tensor.
With this definition, no explicit dimensional splitting is required.
Moreover, we see that in this scheme it is straightforward to
apply both parity-even and parity-odd projectors on the Feynman diagrams:
in fact, parity invariance of the form factors, together with
the block-diagonal form of the projector operators, guarantee that
the parity-even(odd) projectors only need to be applied on the
vector(axial-vector) part of the scattering amplitude.
The definition in~\eqref{eq:larin}, together with the explicit
form of the tensors in~\eqref{eq:proj}, shows that one always has to consider
at most the contraction of one pair of Levi-Civita tensors. The contraction can
be performed assuming that indices are in general $d$-dimensional and this
allows for a straightforward manipulation of the corresponding scalar expressions
in dimensional regularisation.

After acting with the projectors on the Feynman diagrams, it is immediate to manipulate  all resulting Feynman integrals and map them
to integral families. 
At this point, 
our treatment of the Feynman diagrams that only contain massless quarks
proceeds differently compared to those that involve
the exchange of massive top quarks, we therefore discuss the two cases separately below.

\subsection{Massless two-loop corrections}
As already stated in the previous sections, when computing the massless quark contributions, we
consider both a vector and axial-vector interaction and include all relevant Feynman diagrams, 
including pure-singlet and non-singlet diagrams.
While results for the purely vector and for the non-anomalous 
axial-vector part of the amplitude have already been known
in the literature for some time~\cite{Gehrmann:2011ab}, recomputing them  in our current setup 
allows us 
to perform various consistency checks on the calculation, as it will be explained below.

\begin{figure}[t!]
    \centering
    \includegraphics[scale=0.6]{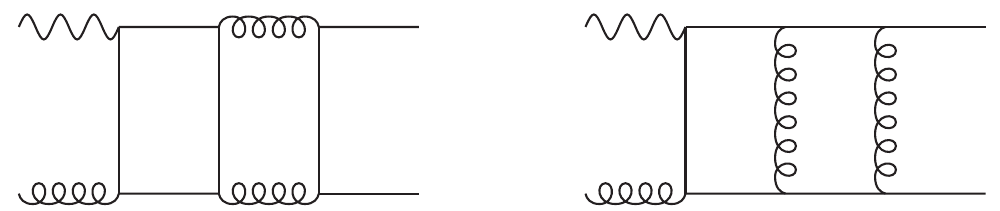}
    \caption{Example of Feynman diagrams for singlet (left) and non-singlet (right) two-loop massless corrections.}
    \label{fig:fig2}
\end{figure}

Up to two loops, all integrals stemming from 
Feynman diagrams which only involve massless virtual quark exchanges
can be easily mapped
to the integral families
originally defined in~\cite{Gehrmann:2000zt,Gehrmann:2001ck}.
Some examples of the relevant Feynman diagrams
are displayed in Figure~\ref{fig:fig2}.

We observe that, after the projectors are applied to the Feynman diagrams, 
only
scalar integrals up to numerator rank 4 appear in the parity-even form factors, 
while the parity-odd
ones require seven-propagator non-planar integrals  up to rank 5.
Despite this, the reduction is relatively simple and can be easily
performed with standard automated codes. We use Reduze 2~\cite{Studerus:2009ye,vonManteuffel:2012np},
which conveniently includes proceedures to map the Feymann diagrams to the 
relevant integral families (and their crossings), and takes care of removing
redundancies among the integrals due to sector symmetries and sector relations, see Ref.~\cite{vonManteuffel:2012np}
for details.
By using the explicit analytic expressions for the master integrals~\cite{Gehrmann:2000zt,Gehrmann:2001ck}, 
it is then straightforward
to obtain analytic expressions for the \emph{bare} 
vector and axial-vector
form factors up to two-loops in terms of so-called
Harmonic Polylogarithms (HPLs)~\cite{Remiddi:1999ew} and 
{2-dimensional Harmonic Polylgarithms} (2dHPLs)~\cite{Gehrmann:2000zt}, which can be evaluated numerically~\cite{Gehrmann:2001pz,Gehrmann:2001jv,Vollinga:2004sn} from their series representations. 

In the modern language of multiple polylogarithms, we define 2dHPLs as 
\begin{align}
    &G(a_1,...,a_n; z) = \int_0^z \frac{dx_1}{x_1-a_1} G(a_2,...,a_n;x_1) \,, \qquad G(0,...,0;z) = \frac{1}{n!} \log^n{z}\,, \nonumber \\
    & G(z) = 1\,, \qquad \mbox{with} \qquad a_i = \left\{ -1,1,0,-y,1-y \right\} \,,
\end{align}
 where for the problem under study, the variables $y,z$ turn out to be any pair of the dimensionless ratios introduced in Eqs.~\eqref{eq:ratios}.
The symmetry of the kinematical constraints allows one to prove that the same set of functions is sufficient
to describe the result for any pair of the variables defined in~\eqref{eq:ratios}. Moreover, it can be shown
that by suitably redefined variables, the same is true in all other kinematically relevant crossings, including
the scattering kinematics~\cite{Gehrmann:2002zr}, allowing to compute
 the amplitudes for $q\bar{q}\to V g$ and $q g \to V q$ in terms of the same class of functions.

\subsection{Massive two-loop corrections}
We compute the leading contributions arising from massive fermion loops to the two-loop pure-singlet axial-vector amplitudes
in the limit of a large top quark mass $m_t\to \infty$.  
While the corresponding one-loop amplitudes start at $\mathcal{O}(1/m_t^2)$, at two loops we also have an $\mathcal{O}(1)$ contribution in that limit.  
While we are mostly interested in this $\mathcal{O}(1)$ contribution, we also include all terms up to $\mathcal{O}(1/m_t^2)$ 
in our results for the pure-singlet axial amplitudes at one and two loops, which allows us to perform consistency checks of  our computational setup and the renormalization and infrared factorization procedure.

\begin{figure}[t!]
    \centering
    \includegraphics[scale=0.6]{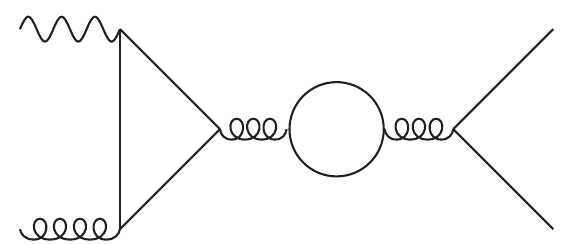}
    \caption{Example of self-energy corrections to the pure-singlet class of diagrams. Both $N_f$ massless and $N_h$ massive quarks
    are allowed to circulate in all fermion loops.}
    \label{fig:fig3}
\end{figure}

We generate the integrands by contracting our twelve projection tensors with the amplitude, keeping the full dependence on the top quark mass. 
As explained above, we limit ourselves to consider mass corrections to 
the subset of diagrams which contribute to the pure-singlet amplitude. 
For consistency, when considering these diagrams we allow massive top quarks also in self-energy corrections, 
as for example in Fig.~\ref{fig:fig3}.
The Feynman integrals in the amplitude are then mapped to two-loop integral families using Reduze~2.

In order to obtain an expansion for large $m_t$, we employ the strategy of regions~\cite{Beneke:1997zp}. 
The top quark mass $m_t$ only appears as an internal mass in a closed sub-loop.  We first focus on two-loop 
diagrams 
with only one massive subloop.  There are two non-vanishing integration regions contributing to the limit $m_t\to\infty$. 
Let $k_1$ and $k_2$ be the two loop momenta.  We identify $k_2$ as the loop momentum running in the fermion loop.  
Hence, the mass $m_t$ only appears in loop propagators depending on $k_2$ (or both $k_1$ and $k_2$) while propagators depending only on $k_1$ are massless.  The first region, which we henceforth call the \emph{large region}, is the one where all loop momenta are large, i.e.\ of the same order as $m_t$,
\begin{equation}\label{eq:regionL}
    k_1 = \mathcal{O}(m_t), \quad k_2 = \mathcal{O}(m_t).
\end{equation}
The second region, which we call the \emph{small region}, is the one where the loop momentum $k_1$ is 
small compared to $m_t$, while $k_2$ is still large,
\begin{equation}\label{eq:regionS}
    k_1 = \mathcal{O}(1), \quad k_2 = \mathcal{O}(m_t).
\end{equation}
The external momenta $p_i$ and the Mandelstam invariants are obviously small, i.e.\ of $\mathcal{O}(1)$, in both regions.  
The expansion of any loop integral is a sum over the two regions.  The contribution of each region is obtained by 
expanding the integrand for $m_t\to\infty$ assuming the loop momenta scale as in \eqref{eq:regionL} for the large region and \eqref{eq:regionS} for the small region.

The expansions in the large region, defined by~\eqref{eq:regionL}, yields vacuum integrals of the form
\begin{equation}
    \int d^{4-2\epsilon} k_1\, d^{4-2\epsilon} k_2\, \frac{1}{(k_1^2)^{a_1}\, (k_2^2-m_t^2)^{a_2}\, ((k_1+k_2)^2-m_t^2)^{a_3}} \label{eq:famL}
\end{equation}
which can all be reduced to products of two one-loop tadpole integrals with mass $m_t$.  In order to map the Laurent expansions of the integrands to integrals belonging to the family defined in~\eqref{eq:famL}, we first need to remove scalar products of the form $k_i\cdot p_j$ from the numerators of the expansions.  This is done via a Passarino-Veltman decomposition~\cite{Passarino:1978jh}, which in this case is very simple since it only depends on the metric tensor $g_{\mu\nu}$ (it can equivalently be performed via an angular integration, see e.g.~\cite{Mastrolia:2016dhn}).

The expansion in the small region, defined by~\eqref{eq:regionS}, yieds instead products of one-loop tadpole integrals with mass $m_t$, times four-point one-loop integrals with massless internal propagators, i.e.\ belonging to the following family
\begin{equation}
    \int d^{4-2\epsilon} k_2\, \frac{1}{(k_2^2-m_t^2)^{a_1}}\, \times \int d^{4-2\epsilon} k_1\, \frac{1}{(k_1)^{2 a_2}\, (k_1+p_1)^{2 a_3}\, (k_1+p_{12})^{2 a_4}\, (k_1+p_{123})^{2 a_5}} \label{eq:famS}
\end{equation}
with $p_{i_1 i_2\cdots}\equiv p_{i_1}+p_{i_2}+\cdots$, or to families obtained from~\eqref{eq:famS} via permutations of external momenta.  The one-loop integrals in $k_1$ are the same integrals that contribute to the one-loop amplitude for diagrams without massive fermions.  Similarly as before, we remove scalar products of the form $k_2\cdot p_j$ and $k_1\cdot k_2$ from the numerators of the expansion via a simple Passarino-Veltman decomposition for the one-loop integral in $k_1$, where only the metric tensor $g_{\mu \nu}$ appears.

We also have a set of diagrams with two massive fermion loops.  These can be cast as products of two one-loop integrals and only contribute to the large region in~\eqref{eq:regionL}, where $k_1$ and $k_2$ are the two loop momenta. After expanding the integrands, these diagrams yield products of two massive one-loop tadpole integrals, which can also be mapped to the family in~\eqref{eq:famL}.

As a check of the consistency of the procedure, we used~\textsc{FiniteFlow}~\cite{Peraro:2016wsq,Peraro:2019svx} to verify, for a selection of integrals, that the expansion by regions commutes with the reduction to master integrals.  In other words, we reconstructed the full $m_t$ dependence of the reduction for a selection of two-loop integrals contributing to the process. We thus performed an expansion by regions of the left-hand sides and the right-hand sides of the reduction identities, verifying that they agree after a further reduction to the master integrals of the families in~\eqref{eq:famL} and~\eqref{eq:famS}.

After reduction to master integrals, the expansion can be expressed in terms of (products of) simple one-loop master integrals, namely one-loop massive tadpole integrals and the massless integrals appearing in the one-loop amplitude.  In particular, all contributions to the pure-singlet axial amplitudes up to $\mathcal{O}(1/m_t^2)$ can be expressed in terms of massless one-loop two-point integrals and one-loop massive tadpoles.

\subsubsection{The unrenormalised form factors up to two loops}
\label{sec:Funren}
Putting everything together, we obtain a result for the 
unrenormalised form factors, including the non-singlet and
pure-singlet corrections up to two loops both for the vector and axial-vector
couplings in massless QCD. Moreover, we also compute the 
axial-vector pure-singlet corrections for massive internal top quarks, keeping the $m_t^0$ and $1/m_t^{2}$ 
terms in an expansion in inverse powers in $m_t$. 
While 
ultimately being interested only in the exact $m_t \to \infty$ limit, keeping the next term in 
the expansion as intermediate step allows us to perform non-trivial checks on our approach.

We  write for the unrenormalised form factors
\begin{align}
\bar{F}_i &= \bar{F}_i^{(0)} + \left( \frac{\bar{\alpha}_s}{2 \pi}\right) \bar{F}_i^{(1)} 
     + \left( \frac{\bar{\alpha}_s}{2 \pi}\right)^2 \bar{F}_i^{(2)} +\mathcal{O}(\bar{\alpha}_s^3)    \nonumber \\
\bar{G}_i &= \bar{G}_i^{(0)} + \left( \frac{\bar{\alpha}_s}{2 \pi}\right) \bar{G}_i^{(1)} 
     + \left( \frac{\bar{\alpha}_s}{2 \pi}\right)^2 \bar{G}_i^{(2)}  + \mathcal{O}(\bar{\alpha}_s^3)\,,      \label{eq:Funren}
    \end{align}
where $\bar{\alpha}_s$ is the bare strong coupling constant. 
We use consistently barred symbols for unrenormlised quantities.

The tree-level contributions to the form factors are not affected by renormalization, such that $F_i^{(0)} = \bar{F}_i^{(0)}$ and 
$G_i^{(0)} = \bar{G}_i^{(0)}$.
The expressions for the tree-level form factors also allow us to 
elaborate upon the statement made
in the context of~\eqref{eq:symtens} on the  
 choice that we made for the tensor multiplying form factor $G_6$. This choice
turns out to be the only one of those described above that is entirely consistent
with the 't Hooft-Veltman prescription. It is only with this
choice that we find that the corresponding tree-level form factors to be 
exactly $d$-independent.
We stress that this is a highly non-trivial
finding, since all algebra at intermediate steps is performed
in $d$ dimensions, see~\eqref{eq:ddimcontr}.
Explicitly, for the vector tree-level contribution
we find
\begin{align}
   & F_1^{(0)} = 0\,, \quad
    F_2^{(0)} = -\frac{2 \left(s_{13}+s_{23}\right)}{s_{12} s_{13} s_{23}}\,,\quad
    F_3^{(0)} =\frac{s_{13}-s_{23}}{s_{13} s_{23}}
    \,, \nonumber \\
     & F_4^{(0)} = \frac{-s_{23}^2-s_{12} s_{23}-s_{13} s_{23}+s_{12} s_{13}}{s_{12} s_{13} s_{23}}\,, \quad  F_5^{(0)} = 
    \frac{s_{13}^2+s_{12} s_{13}+s_{23} s_{13}-s_{12} s_{23}}{s_{12} s_{13} s_{23}}\,, \nonumber  \\
   &
    F_6^{(0)} = \frac{2 s_{12}+s_{13}+s_{23}}{s_{12} s_{13}}
\end{align}

Similarly, for the axial-vector tree-level (which is only of non-singlet type)
contribution we have 

\begin{align}
G_1^{(0)} &= -\frac{-s_{23}^2-s_{12} s_{23}-3 s_{13} s_{23}+s_{12} s_{13}}{s_{12} s_{13}^2 s_{23}^2}\,,\quad
G_2^{(0)} = -\frac{s_{13}^2+s_{12} s_{13}+3 s_{23} s_{13}-s_{12} s_{23}}{s_{12} s_{13}^2 s_{23}^2}\,, \nonumber \\
G_3^{(0)} &= -\frac{4 s_{12}+s_{13}+3 s_{23}}{s_{12} s_{13}^2 s_{23}}\,, \quad
G_4^{(0)} = -\frac{2 \left(2 s_{12}^2+s_{13} s_{12}+3 s_{23} s_{12}+s_{23}^2-s_{13}
   s_{23}\right)}{s_{12}^2 s_{13}^2 s_{23}}\,, \nonumber \\
G_5^{(0)} &= -\frac{2 \left(2 s_{12}^2+3 s_{13} s_{12}+s_{23} s_{12}+s_{13}^2-s_{13}
   s_{23}\right)}{s_{12}^2 s_{13}^2 s_{23}}\,, 
\quad
G_6^{(0)} = -\frac{s_{13}-s_{23}}{s_{12} s_{13} s_{23}}\,. \label{eq:treelevG}
\end{align}

All form factors receive non-singlet contributions, which we denote by $\bar{F}_i^{(0,1,2),n}$  and $\bar{G}_i^{(0,1,2),n}$ . The pure-singlet contributions only enter 
at two loops in the vector form factors, denoted by $\bar{F}_i^{(2),p}$, due to a generalization of Furry's theorem~\cite{Furry:1937zz}, and at one and two loops in the axial-vector form factors, denoted by  $\bar{G}_i^{(1,2),p}$. 
Since pure-singlet and non-singlet contributions will be dressed with different electroweak charge factors, they are treated separately throughout.


\section{UV renormalisation and subtraction of the IR poles}
\label{sec:renorm}
The unrenormalised form factors (\ref{eq:Funren}) contain in general 
divergences both of ultraviolet (UV) and  infrared (IR) origin, which 
manifest as 
poles in the dimensional regulator paramenter $\epsilon$.

UV poles can be consistently removed by the procedure of renormalisation.
In our case, renormalisation proceeds  differently in the purely vector and axial-vector cases and in the non-singlet and pure-singlet contributions, for two different reasons.
First of all, as already discussed,
due to the axial anomaly the pure-singlet axial-vector contribution can only be 
computed consistently in the full Standard Model if the closed quark loop is summed over a complete isospin doublet.
 A well-defined result for the pure-singlet axial-vector contribution with a single quark flavour in the loop requires an 
 extra renormalization of the axial-vector pure-singlet current~\cite{Larin:1993tq}. 
This pure-singlet renormalization of the axial anomaly is required for any $\gamma_5$ scheme. Secondly, since we work in the Larin scheme, the  non-singlet axial-vector
coupling also requires a renormalization
to obtain consistent results~\cite{Larin:1993tq}.

We ultimately include top-quark effects in the exact  
$m_t \to \infty$ limit, 
but also compute corrections up to order $1/m_t^2$ in order
to consistently check our renormalisation procedure. The formulae below
reflect therefore the full renormalisation procedure required up to
this order in $1/m_t$.

According to the discussion above, we renormalise $\alpha_s$ by replacing
the bare coupling constant as follows
\begin{align}
    \bar{\alpha}_s \mu_0^{2 \epsilon} S_\epsilon 
    = \alpha_s \mu_R^{2 \epsilon}\left[ 1 
    - \frac{1}{\epsilon} \left( \beta_0 + N_h\, \delta_w \right)
    \left( \frac{\alpha_s}{2 \pi} \right) + 
    \left( \frac{\beta_0^2}{\epsilon^2} - \frac{\beta_1}{2 \epsilon}\right)
    \left( \frac{\alpha_s}{2 \pi} \right)^2 + \mathcal{O}(\alpha_s^3)
    \right]
\end{align}
where $S_\epsilon = (4 \pi)^\epsilon e^{-\epsilon \gamma_E}$,
with $\gamma_E = 0.5772...$ the Euler-Mascheroni constant, 
\begin{equation}
    \delta_w = -\frac{2}{3} T_R \left( \frac{m_t^2}{\mu_R^2}\right)^{-\epsilon}\,.
\end{equation} 
We also introduced $N_h$ for the number of heavy flavours of mass $m_t$, such that $N_h=1$ for the pure-singlet axial-vector contribution and zero otherwise. We use the tag $N_h$ to compactly include the dependence on the top mass in the beta function only for the pure-singlet axial-vector contribution, which 
in this calculation is required only
to first order in $\alpha_s$ to renormalise the two-loop amplitude.
$\beta_i$ are the coefficients of the massless  QCD beta function and read
\begin{align}
    \beta_0 = \frac{11}{6}C_A - \frac{4}{6} T_R N_f\,, \qquad 
    \beta_1 = \frac{17}{6}C_A^2 - \frac{10}{6} T_R C_A N_f - T_R C_F N_f\,,
\end{align}
where for $SU(N)$ we have $$C_A = N\,, \quad C_F = \frac{N^2-1}{2 N}\,, \quad
T_R = \frac{1}{2}\,,$$
with $N$ the number of colors and $N_f$ the number of active massless
flavours. We must keep in mind that an overall factor of $\bar{\alpha}_s^{1/2}$ is contained in 
the relation  (\ref{eq:ampdef}) between amplitudes and form factors. For simplicity, we set $\mu_0^2 = \mu_R^2=s_{12}+s_{13}+s_{23} = q^2$ throughout.

In addition to this, the renormalisation of the non-singlet axial-vector
current requires multiplying the non-singlet contributions to the axial-vector form factors $G_i$
by the non-singlet renormalisation constant, whose explicit value depends on the
scheme used for treating $\gamma_5$. In the Larin scheme, the latter is known up to four loops~\cite{Chen:2021gxv}. We only require its value up to two loops in our calculation: 
\begin{align}
    Z_a^{n} = 1 -2 C_F \left( \frac{\alpha_s}{2 \pi} \right)
    + \left[ \frac{11}{2} C_F^2 - \frac{107}{36} C_F C_A
    + \frac{1}{18} C_F N_f \right]\left( \frac{\alpha_s}{2 \pi} \right)^2
    + \mathcal{O}(\alpha_s^3)
\end{align}

The renormalization of the axial-vector current is usually stated~\cite{Larin:1993tq}
 in the form of renormalization constants for the non-singlet and the singlet currents, with the singlet current being the sum of non-singlet and pure-singlet. 
 We choose instead to reformulate the renormalization in terms of non-singlet and pure-singlet, as introduced in~\cite{Gehrmann:2021ahy}. In this form,
 the currents are mapped more
 easily onto the electroweak coupling factors, but at the expense of a mixing of non-singlet into pure-singlet contributions under renormalization. The pure-singlet 
 renormalization constant starts only at order $\alpha_s^2$ and reads~\cite{Larin:1993tq} for a single quark flavour:
 \begin{align}
    Z_a^{s} = C_FT_R \left( \frac{\alpha_s}{2 \pi} \right)^2
      \left[ \frac{3}{2\epsilon} + \frac{3}{4} \right] 
    + \mathcal{O}(\alpha_s^3)\,, \label{eq:zax}
\end{align}
where the pole part stems from the UV renormalization of the axial anomaly and the finite part is required to restore the axial Ward identities in dimensional regularization in the 
Larin scheme.

Finally, for what concerns  the  pure-singlet contributions from virtual top quarks to the axial-vector form factors, 
we also renormalise $m_t$ in the on-shell scheme
\begin{align}
    \bar{m}_{t} = m_t \left[ 1 + \left( \frac{\alpha_s}{2 \pi} \right) \delta_m \right] + \mathcal{O}(\alpha_s^2)
\end{align}
where $\bar{m}_t$ is the bare mass and
\begin{equation}
\delta_m = C_F \left( \frac{m_t^2}{\mu_R}\right)^{-\epsilon}\left( -\frac{3}{2 \epsilon} -2 + \mathcal{O}(\epsilon)\right)\,.
\end{equation}
Finally, for these contributions gluon wave function renormalisation is performed by multiplying the 
pure-singlet axial-vector part of the form factors with
\begin{equation}
    Z_A^{1/2} = 1 + \frac{1}{2}\left( \frac{\alpha_s}{2 \pi} \right) N_h \delta_w
    + \mathcal{O}(\alpha_s^2)
\end{equation}
since there is only one external gluon.

The UV-renormalized form factors are expanded in the renormalized coupling constant $\alpha_s = \alpha_s (\mu^2)$:
\begin{align}
{F}_i &= {F}_i^{(0)} + \left( \frac{\alpha_s}{2 \pi}\right) {F}_i^{(1)} 
     + \left( \frac{\alpha_s}{2 \pi}\right)^2 {F}_i^{(2)} +\mathcal{O}(\alpha_s^3) \,,   \nonumber \\
{G}_i &= {G}_i^{(0)} + \left( \frac{\alpha_s}{2 \pi}\right) {G}_i^{(1)} 
     + \left( \frac{\alpha_s}{2 \pi}\right)^2 {G}_i^{(2)}  + \mathcal{O}(\alpha_s^3)\,.     \label{eq:Fren}
    \end{align}
We again distinguish non-singlet and pure-singlet contributions by a superscript $(n,p)$.

Putting everything together, we obtain renormalised form factors for the various contributions.
For the vector form factors, we find for non-singlet and pure-singlet contributions:
\begin{align}
F_i^{(0),n} & = \bar{F}_i^{(0),n}\,, \nonumber \\
F_i^{(1),n}   &= S_\epsilon^{-1} \bar{F}_i^{(1),n} 
    - \frac{\beta_0}{2 \epsilon}  \bar{F}_i^{(0),n}\,, \nonumber \\
   F_i^{(2),n}   &= S_\epsilon^{-2} \bar{F}_i^{(2),n}  
    - \frac{3\beta_0}{2 \epsilon}  \bar{F}_i^{(1),n}  S_\epsilon^{-1} 
    - \left( \frac{\beta_1}{4 \epsilon} - \frac{3\beta_0^2}{8 \epsilon^2}\right) F_i^{(0),n}\,,  \label{eq:ffrenorm} \\
  F_i^{(2),p} & = \bar{F}_i^{(2),p}  \,,
\end{align}
while for the axial-vector non-singlet part we have 
\begin{align}
G_i^{(0),n}  = &\; \bar{G}_i^{(0),n}   \,, \nonumber \\
   G_i^{(1),n}   =&\; S_\epsilon^{-1}  \bar{G}_i^{(1),n} 
    - \frac{\beta_0}{2 \epsilon}  \bar{G}_i^{(0),n} 
     -  2 C_F \bar{G}_i^{(0),n}\,, \nonumber \\
   G_i^{(2),n}   =&\; S_\epsilon^{-2} \bar{G}_i^{(2),n} 
    - \frac{3\beta_0}{2 \epsilon}  \bar{G}_i^{(1),n}   S_\epsilon^{-1} 
    - \left( \frac{\beta_1}{4 \epsilon} - \frac{3\beta_0^2}{8 \epsilon^2}\right)  \bar{G}_i^{(0),n}    \nonumber \\
     &-2 C_F   \bar{G}_i^{(1),n}   S_\epsilon^{-1}
     + C_F \left( \frac{2 \beta_0}{\epsilon} + \frac{11}{2}C_F
     - \frac{107}{36}  C_A
             + \frac{1}{18} N_f\right)  \bar{G}_i^{(0),n} \,,
\end{align}
where we stress that the extra terms  in comparison to 
(\ref{eq:ffrenorm})  stem from the
 renormalisation of the axial-vector current in Larin's scheme.

For the pure-singlet axial-vector form factors, we consider the contributions from virtual massless and virtual massive quarks separately. We assemble the resulting 
one-loop and 
two-loop pure-singlet contributions to the form factors depending on whether the axial-vector current couples to a massless quark $(\bar{G}_i^{(j),p0})$ or to a massive quark 
 $(\bar{G}_i^{(j),pm})$ of mass $m_t$. It should be noted that both these contributions contain massive and massless quark bubble insertions into their 
 gluon propagators and external gluon legs.

 With this in mind, the renormalization of the pure-singlet axial-vector form factors reads:
  \begin{align}
 G_i^{(1),p0}  =&\;  S_\epsilon^{-1} \bar{G}_i^{(1),p0}   \,, \nonumber \\
 G_i^{(1),pm}  =&\; S_\epsilon^{-1} \bar{G}_i^{(1),pm}   \,, \nonumber \\
G_i^{(2),p0} = &\; S_\epsilon^{-2}  \bar{G}_i^{(2),p0} 
    - \left( \frac{ 3\beta_0}{2 \epsilon} + \frac{N_h \delta_w}{\epsilon} \right) S_\epsilon^{-1} \bar{G}_i^{(1),p0} 
     + C_FT_R  \left( \frac{3}{2\epsilon} + \frac{3}{4} \right)  \bar{G}_i^{(0),n} \nonumber \\
     &\; -2 C_F   \bar{G}_i^{(1),p0}   S_\epsilon^{-1}  \, ,  \nonumber \\
  G_i^{(2),pm} =&\; S_\epsilon^{-2}  \bar{G}_i^{(2),pm} 
    - \left( \frac{3\beta_0}{2 \epsilon} + \frac{N_h \delta_w}{\epsilon} \right) S_\epsilon^{-1} \bar{G}_i^{(1),pm} 
     + S_\epsilon^{-1}\,\frac{d\bar{G}_i^{(1),pm}}{dm_t}\,\delta_m \nonumber \\ 
     &\; + C_FT_R  \left( \frac{3}{2\epsilon} + \frac{3}{4} \right)  \bar{G}_i^{(0),n} \, 
     -2 C_F   \bar{G}_i^{(1),pm}   S_\epsilon^{-1} , 
 \end{align}
where the next-to-last term on the right hand sides of the last two equations comes from the renormalization (\ref{eq:zax}) of the axial anomaly.  
These terms cancel out in the difference $G_i^{(2),pm}- G_i^{(2),p0}$, since the summation over an isospin doublet renders the theory anomaly-free. We stress also that the last term is a manifestation of the mixing of non-singlet and pure-singlet currents under renormalization.

The renormalised form factors still have poles of IR nature.
IR singularities in QCD are universal and their structure only depends on the 
number and type
of strongly interacting partons involved in the scattering process~\cite{Catani:1998bh,Becher:2009qa,Dixon:2009ur}.
For the case under study, we write the finite remainders 
in terms of the renormalised
coefficients as:
\begin{eqnarray}
F_{i,\fin}^{(1),n} &=&  F_{i}^{(1),n}  -  I_1(\epsilon) F_{i}^{(0),n}\,, \nonumber \\
F_{i,\fin}^{(2),n} &=& F_{i}^{(2),n}  - I_1(\epsilon) F_{i}^{(1),n} 
    - I_2(\epsilon) F_{i}^{(0),n} \,, \nonumber \\
    G_{i,\fin}^{(1),n} &=&  G_{i}^{(1),n}  -  I_1(\epsilon) G_{i}^{(0),n}\,, \nonumber \\
G_{i,\fin}^{(2),n} &=& G_{i}^{(2),n}  - I_1(\epsilon) G_{i}^{(1),n} 
    - I_2(\epsilon) G_{i}^{(0),n} \,, 
\label{eq:nsfinite}
\end{eqnarray}
and
\begin{eqnarray}
G_{i,\fin}^{(2),p0} &=&  G_{i}^{(2),p0}  -  I_1(\epsilon) G_{i}^{(1),p0}\,, \nonumber \\
G_{i,\fin}^{(2),pm} &=&  G_{i}^{(2),pm}  -  I_1(\epsilon) G_{i}^{(1),pm}\,.
\label{eq:psfinite}
\end{eqnarray}
We note that $F_i^{(2),p}$, $G_i^{(1),p0}$ and $G_i^{(1),pm}$ are already infrared-finite.

For our process, since we only have three coloured particles, 
the Catani operators~\cite{Catani:1998bh} $I_1(\epsilon)$ and $I_2(\epsilon)$ are 
diagonal in color space. The operator $I_1$ reads
\begin{align}
    I_1(\epsilon) = - \frac{e^{\epsilon \gamma_E}}{2\Gamma(1-\epsilon)}
    \left[ N \left( \frac{1}{\epsilon^2} + \frac{3}{4\epsilon} 
          + \frac{\beta_0}{2 N \epsilon}\right) 
          \left( {\tt S}_{13} + {\tt S}_{23} \right)
          - \frac{1}{N} \left( \frac{1}{\epsilon^2} + \frac{3}{2\epsilon} \right) 
          {\tt S}_{12}\right]\,,
\end{align}
with 
\begin{equation}
    {\tt S}_{ij} = \left( -\frac{s_{ij}}{q^2}\right)^{-\epsilon}\,.
\end{equation}
The operator $I_2(\epsilon)$ instead reads
\begin{align}
    I_2(\epsilon) &=  -\frac{1}{2} I_1(\epsilon)^2 - \frac{\beta_0}{\epsilon}I_1(\epsilon)
    + e^{-\epsilon \gamma_E}\frac{\Gamma(1-2 \epsilon)}{ \Gamma(1-\epsilon)}
    \left(\frac{\beta_0}{\epsilon} + K \right)I_1(2\epsilon) + H_2(\epsilon)\,,
\end{align}
where
\begin{align}
    H_2(\epsilon) = \frac{e^{\epsilon \gamma_E}}{4 \epsilon \, \Gamma(1-\epsilon)}
    {\tt H}_2\,,
\end{align}
with
\begin{eqnarray}
\label{eq:Htwo}
{\tt H}_2 &=&  
\left(4\zeta_3+\frac{589}{432}- \frac{11\pi^2}{72}\right)N^2
+\left(-\frac{1}{2}\zeta_3-\frac{41}{54}-\frac{\pi^2}{48} \right)
+\left(-3\zeta_3 -\frac{3}{16} + \frac{\pi^2}{4}\right) \frac{1}{N^2}\nonumber \\
&&
+\left(-\frac{19}{18}+\frac{\pi^2}{36} \right) N N_f 
+\left(-\frac{1}{54}-\frac{\pi^2}{24}\right) \frac{N_f}{N}+ \frac{5}{27} N_f^2.
\end{eqnarray}
and finally the constant $K$ is
\begin{equation}
K = \left( \frac{67}{18} - \frac{\pi^2}{6} \right) C_A - 
\frac{10}{9} T_R N_f.
\end{equation}
In all the above expressions for the Catani operators, 
$N_f$ is taken equal to the number of light quark flavours in all non-singlet contributions, and equal to the number of light quark flavours plus one 
in the pure-singlet contributions.

The finiteness of (\ref{eq:nsfinite}) was already established~\cite{Garland:2002ak} for the vector form factors. The finiteness of the axial-vector form factors, especially 
in the pure-singlet case (\ref{eq:psfinite}), where it was also checked at order $1/m_t^2$ in the large-mass expansion, and which 
takes place prior to summation over an isospin doublet,  represents a strong check on the internal consistency of our calculation.


\section{The helicity amplitudes}
\label{sec:helamp}

The form factors that 
were computed in the previous section describe the 
coupling of an external off-shell vector 
or axial-vector current to a $q\bar q g$-system. 
From these 
form factors, it is possible 
to reconstruct the helicity amplitudes for the decay of 
any vector boson $V=\gamma^*,Z,W^\pm$ 
into $q\bar q g$ by multiplying them
by the appropriate electroweak couplings. 
Helicity amplitudes for vector-boson-plus-parton production 
can then be obtained by analytical continuation to 
the appropriately crossed kinematical regions~\cite{Gehrmann:2002zr}. 

\subsection{Helicity amplitudes for an external vector and axial-vector current}
It is useful to start off by considering the helicity amplitudes for the case of a 
generic vector or axial-vector current.
We use the spinor helicity formalism and let the off-shell current decay to two massless leptons
\begin{equation}
   V(p_4) \to l(p_5) + \bar{l}(p_6)  \,.
\end{equation}
We call $\lambda_{f}$ the helicity of fermion $f$ and $\lambda_3$  the helicity of the external gluon $g(p_3)$.
We define the left- and right-handed currents 
for a pair of fermions as
\begin{equation}
C_{L}^\mu(p,q) =  \bar{u}_L(q) \gamma^\mu u_L(p) = \langle q \gamma^\mu p ], \qquad
C_{R}^\mu(p,q) = \bar{u}_R(q) \gamma^\mu u_R(p) = [ q \gamma^\mu p \rangle.
\end{equation}
The polarization vector for the external gluon, with positive and negative helicity respectively, is instead given by (remembering that we picked $p_2$ as gauge vector, and that the helicities are defined for $p_3$ outgoing)
\begin{equation}
\epsilon_{3,-}^\mu = \frac{\langle 3 \gamma^\mu 2 ]}{\sqrt{2} [32]}\,, \quad 
\epsilon_{3,+}^\mu = \frac{\langle 2 \gamma^\mu 3 ]}{\sqrt{2} \langle 23 \rangle}\,.
\end{equation}
Finally, in order to write down the helicity amplitudes
starting from our general tensor structure in ~\eqref{eq:AAV},
we  also use the following four-dimensional representation of the vector $v_A^\mu$
\begin{equation}
v_A^\mu \equiv \epsilon^{p_1 p_2 p_3 \mu} = \frac{1}{4} \Big[ [ 1 2 3 \gamma^\mu 1\rangle -\langle 1 2 3 \gamma^\mu 1] \Big].
\end{equation}
Note that, at this stage, $v_A$ will only be contracted with four-dimensional external states, hence its four-dimensional representation is sufficient here, although $d$-dimensional identities were used in  computing the form factors.

With these definitions, we consider the quantity
\begin{align}
{\rm M}_{\lambda_{q_2}\lambda_3 \lambda_{l_5}} = \eps_{3,\rho}^{\lambda_3}   \, 
	A^{\mu \rho}_{\lambda_{\bar q_1}\lambda_{q_2} } C_{\lambda_{l_5}}^\mu(p_5,p_6)\,,
	 \label{eq:HAnocouplings}
\end{align}
where $A^{\mu \rho}_{\lambda_{\bar q_1}\lambda_{q_2} } $ is obtained from the general decomposition for the amplitude in~\eqref{eq:AAV} by fixing quark and gluon helicities. We stress that this means that we assign helicities according to the handedness of the outgoing fermions.
We consider for definiteness the case of left-handed quark and lepton currents.
By considering the vector and axial-vector current separately, 
we can write for both in spinor helicity formalism as
\begin{align}
    {\rm M}_{L+L}^{v} = \frac{1}{\sqrt{2}} \, 
	 \bigg[ \langle 1 2 \rangle [1 3]^2 \Big( \alpha_1 \langle 536] 
	+ \alpha_2 \langle 526] \Big)
        + \alpha_3 \langle 25 \rangle [13] [36] \bigg]\,,
\label{eq:mLLv}
\end{align}
\begin{align}
    {\rm M}_{L+L}^{a} = \frac{1}{\sqrt{2}} \, 
	 \bigg[ \langle 1 2 \rangle [1 3]^2 \Big( \beta_1 \langle 536] 
	+ \beta_2 \langle 526] \Big) 
        + \beta_3 \langle 25 \rangle [13] [36] \bigg] \,,
\label{eq:mLLa}
\end{align}
where the superscrpt $(v,a)$ indicate the vector and axial-vector parts. 
The coefficients $\alpha_i$ encompass the contribution from the former while the $\beta_i$ indicate the latter. 
They can be written in terms of the original form factors as
\begin{align}
  \alpha_1 = - F_1\,, \quad
  \alpha_2 = F_2-F_1 + \frac{2 F_6}{s_{23}} \,,
  \quad
  \alpha_3 = 2 F_3 - \frac{2 s_{12} F_6}{s_{23}}\,,
  \label{eq:LmLv}
\end{align}
\begin{align}
    \beta_1 &= \frac{1}{2} \Big[s_{23} \left(G_1 + G_3\right) + s_{12} \left(G_3 - G_4\right)\Big]  \, ,\nonumber \\ 
    \beta_2 &= \frac{1}{2} \Big[s_{13} G_3 + s_{23} \left(G_1 - G_2 + G_3\right) - s_{12} G_4 + s_{12} G_5 - 
   2 G_6\Big]\, ,\nonumber \\
    \beta_3 &= s_{12} \left(G_6 -s_{13} G_3\right)\,.
\label{eq:LmLa}    
\end{align}
We stress that both the vector and the axial-vector part can be decomposed in terms of the same
spinor structures.

Similarly, for the opposite choice of the gluon helicity we find
\begin{align}
	 {\rm M}^v_{L-L} &=  \frac{1}{\sqrt{2}} \, 
	 \bigg[ \langle 23 \rangle^2 [1 2] \Big( \gamma_1 \langle 536] 
	+ \gamma_2 \langle 516] \Big)
        + \gamma_3 \langle 23 \rangle \langle 35 \rangle [16] \bigg]\,, \label{eq:LpLv}
\end{align}
\begin{align}
	 {\rm M}^a_{L-L} &=  \frac{1}{\sqrt{2}} \, 
	 \bigg[ \langle 23 \rangle^2 [1 2] \Big( \delta_1 \langle 536] 
	+ \delta_2 \langle 516] \Big)
        + \delta_3 \langle 23 \rangle \langle 35 \rangle [16] \bigg]\,, \label{eq:LpLa}
\end{align}
where
\begin{align}
    \gamma_1 &= \frac{1}{s_{23}} \Big[ s_{13} F_2 + 2 \left(F_5-F_3\right)\Big] \,, 
    \nonumber \\
    \gamma_2 &= \frac{1}{s_{23}}
    \Big[s_{13}(F_2-F_1) - 2(F_4-F_5+F_6)\Big]\,,
    \nonumber \\
    \gamma_3 &= - 2\frac{(s_{23} F_3 + s_{12} F_6)}{s_{23}}\,,
\end{align}
\begin{align}
    \delta_1 &=\frac{1}{2 s_{23}} \Big[s_{13} (s_{23} G_2 - (s_{12} + s_{13}) G_3 + s_{12} G_5) - 2 (s_{12} + s_{13}) G_6 \Big]\,, 
    \nonumber \\
    \delta_2 &= -\frac{1}{2 s_{23}}\Big[s_{13}^2 G_3 + 4 s_{23} G_6 + 
 s_{13} (s_{23} (G_1 - G_2 + G_3) + s_{12} G_4 - s_{12} G_5 + 2 G_6)\Big]\,, \nonumber \\
    \delta_3 &= -s_{12}  \left(s_{13} G_3 + 3 G_6\right)\,.
\end{align}

Similar relations can be written for a right-handed quark current, which are related to the ones above by parity conjugation, with a relative minus sign between vector and axial results,
\begin{align}
	 {\rm M}^v_{R+L} &=  -\frac{1}{\sqrt{2}} \, 
	 \bigg[ [23]^2 \langle 1 2\rangle \Big( \gamma_1 \langle 536] 
	+ \gamma_2 \langle 516] \Big)
        + \gamma_3 [23] [36]  \langle 15 \rangle \bigg]\,, 
\label{eq:mRLv1}
\end{align}
\begin{align}
	 {\rm M}^a_{R+L} &=  \frac{1}{\sqrt{2}} \, 
	 \bigg[ [23]^2 \langle 1 2\rangle \Big( \delta_1 \langle 536] 
	+ \delta_2 \langle 516] \Big)
        + \delta_3 [23] [36]  \langle 15 \rangle \bigg]\,, 
\end{align}
\begin{align}
    {\rm M}_{R-L}^{v} = -\frac{1}{\sqrt{2}} \, 
	 \bigg[ [1 2] \langle 1 3\rangle^2 \Big( \alpha_1 \langle 536] 
	+ \alpha_2 \langle 526] \Big)
        + \alpha_3 [26] \langle 13\rangle \langle 35\rangle \bigg]\,,
\end{align}
\begin{align}
    {\rm M}_{R-L}^{a} = \frac{1}{\sqrt{2}} \, 
	 \bigg[ [1 2] \langle 1 3\rangle^2 \Big( \beta_1 \langle 536] 
	+ \beta_2 \langle 526] \Big)
        + \beta_3 [26] \langle 13\rangle \langle 35\rangle \bigg] \,,
\label{eq:mRLa2}
\end{align}
where $\alpha_i$, $\beta_i$, $\gamma_i$ and $\delta_i$ are the same given above in terms of the form factors.

The finite remainder for vector and axial-vector helicity coefficients 
$\Omega_i = \left\{ \alpha_i,\gamma_i \right\}$ and 
$\Lambda_i = \left\{\beta_i, \delta \right\}$ can be expanded as series in the strong coupling
\begin{align}
\Omega_i &= \Omega_i^{(0)} + \left( \frac{\alpha_s}{2 \pi}\right) \Omega_i^{(1)} 
     + \left( \frac{\alpha_s}{2 \pi}\right)^2 \Omega_i^{(2)} +\mathcal{O}(\alpha_s^3)\,, \nonumber \\
\Lambda_i &= \Lambda_i^{(0)} + \left( \frac{\alpha_s}{2 \pi}\right) \Lambda_i^{(1)} 
     + \left( \frac{\alpha_s}{2 \pi}\right)^2 \Lambda_i^{(2)} +\mathcal{O}(\alpha_s^3)\,.  \label{eq:OmLam}   
\end{align}
At $l$-loops, they inherit the corresponding decomposition into
non-singlet and pure-singlet from the finite remainder of the form factors computed in the previous section. 
Moreover for the pure-singlet axial-vector case, we split them further into
massless and massive contributions.
We indicate the various components as above with $\Omega_i^{(l),n}$, $\Omega_i^{(l),p}$,
$\Lambda_i^{(l),n}$, $\Lambda_i^{(l),p0}$, $\Lambda_i^{(l),pm}$ respectively.

At this point, we can perform an important consistency check on our calculation.
We verified that after UV renormalisation and IR subtraction, as expected, the non-singlet contribution for the vector and axial-vector helicity amplitudes agree to the two-loop order. We find, in particular
\begin{align}
&\alpha_i^{(0)} = - \beta_i^{(0)}\, \quad \gamma_i^{(0)} = - \delta_i^{(0)}\,, \nonumber \\
&\alpha_i^{(1),n} = - \beta_i^{(1),n}\,, \quad \gamma_i^{(1),n} = - \delta_i^{(1),n}\,, \nonumber \\
&\alpha_i^{(2),n} = - \beta_i^{(2),n}\,, \quad \gamma_i^{(2),n} = - \delta_i^{(2),n}\,.
\label{eq:NSavv}
\end{align}
The minus-sign between the $\alpha,\gamma$ (vector) and 
$\beta,\delta$ (axial-vector) coefficients arises from
the definition of the basic vector and axial-vector 
amplitudes in terms of left-handed (V-A) 
quark currents in (\ref{eq:mLLv},\ref{eq:mLLa}), which 
are then compensated by the explicit minus-signs 
between vector and axial-vector amplitudes for 
right-handed (V+A) quark currents 
in (\ref{eq:mRLv1}--\ref{eq:mRLa2}).

As exemplification of our results, we conclude this section by providing some analytic formulas for the axial-vector
pure-singlet contributions.
At tree level there is no pure-singlet contribution, so  we provide the 
non-singlet results, which read
\begin{align}
\beta_1^{(0)} = 0\,, \quad \beta_2^{(0)} = -\frac{4}{q^4} \frac{1}{yz}\,,
\quad \beta_3^{(0)} = \frac{4}{q^2} \frac{1-y}{yz} \,,
\end{align}
\begin{align}
\delta_1^{(0)} = 0\,, \quad \delta_2^{(0)} = \frac{4}{q^4} \frac{1}{yz}\,,
\quad \delta_3^{(0)} = \frac{4}{q^2} \frac{1-z}{yz}\,.
\end{align}
At one loop instead we find for the pure-singlet contributions
\begin{align}
&\beta_1^{(1),pm} = \beta_2^{(1),pm} = \beta_3^{(1),pm} = \mathcal{O}\left( \frac{1}{m_t^2} \right)\,,\nonumber \\
&\delta_1^{(1),pm} = \delta_2^{(1),pm} = \delta_3^{(1),pm} = \mathcal{O}\left( \frac{1}{m_t^2} \right)\,,
\end{align}
\begin{align}
&\beta_1^{(1),p0} = 0\,, \quad \beta_2^{(1),p0} = 0\,,\quad\beta_3^{(1),p0} = \frac{4}{q^2} \left[ \frac{1}{y+z} + \frac{\log{(1-y-z)}}{(y+z)^2}   \right]\,,
\end{align}
\begin{align}
&\delta_1^{(1),p0} = 0\,, \quad \delta_2^{(1),p0} = 0\,, \quad \delta_3^{(1),p0} = \frac{4}{q^2} \left[ \frac{1}{y+z} + \frac{\log{(1-y-z)}}{(y+z)^2}   \right]\,,
\end{align}
where we notice that, only at one-loop order, $\delta_i^{(1),p0} = \beta_i^{(1),p0}$ for $i=1,2,3$.

Finally, at two loops the pure-singlet coefficients
corresponding to the axial-vector current coupled to a massive quark loop
are very compact. Limiting ourselves to the order zero in the large-mass expansion we find
\begin{align}
&\beta_1^{(2),pm} = \mathcal{O}\left( \frac{1}{m_t^2} \right)\,, \quad 
\beta_2^{(2),pm} = -\frac{3 C_F}{q^4} \frac{1}{yz} \left[ 1 +2 \log{\frac{m_t^2}{q^2}} \right]
+ \mathcal{O}\left( \frac{1}{m_t^2} \right)\,, 
\nonumber \\
&\beta_3^{(2),pm} = \frac{3 C_F}{q^2} \frac{1-y}{yz} \left[ 1 +2 \log{\frac{m_t^2}{q^2}} \right]
+ \mathcal{O}\left( \frac{1}{m_t^2} \right)\,,
\end{align}
\begin{align}
&\delta_1^{(2),pm} = \mathcal{O}\left( \frac{1}{m_t^2} \right)\,, \quad 
\delta_2^{(2),pm} = \frac{3 C_F}{q^4} \frac{1}{yz} \left[ 1 +2 \log{\frac{m_t^2}{q^2}} \right]
+ \mathcal{O}\left( \frac{1}{m_t^2} \right)\,, 
\nonumber \\
&\delta_3^{(2),pm} = \frac{3 C_F}{q^2} \frac{1-z}{yz} \left[ 1 +2 \log{\frac{m_t^2}{q^2}} \right]
+ \mathcal{O}\left( \frac{1}{m_t^2} \right)\,.
\end{align}
The corresponding results for the coupling to a massless quark loop are lenghtier and can be obtained from the ancillary files in electronic format. As an example, we provide here the result for the $\beta_1^{(2),p0}$.
We decompose it according to  different powers of $N$ as follows
\begin{align}
\beta_1^{(2),p0} &= \frac{1}{q^4} \left[ N K_1 + \frac{1}{N} K_2 \right]\,,
\end{align}
where the functions $K_1$ and $K_2$ are linear combinations of 2-dimensional
harmonic polylogarithms with rational coefficients.
In particular we find for the first 
$$K_1 = \sum_{i=1}^{13} R_i Q_i\,,$$ with
\allowdisplaybreaks
\begin{align}
& R_1=\frac{1}{y^3}\,, \quad
 R_2=\frac{1}{y (z-1)} \,, \quad
 R_3=\frac{z}{(z-1) (y+z-1)^2} \,, \quad
 R_4=\frac{z}{y^2 (y+z)^2} \,,\nonumber \nonumber\\
& R_5=\frac{y-z+2}{y (z-1) (y+z)} \,, \quad
 R_6=\frac{3 y+2 z}{y^2 (y+z)^2} \,, \quad
 R_7=\frac{5 y^2+5 y z+4 z-4}{y (z-1) (y+z-1) (y+z)}\,, \nonumber \\
& R_8=\frac{y^2+6 y z+3 z^2}{y^2 z (y+z)^2} \,, \quad
 R_9=\frac{5 y^2-2 y z+2 y+4 z^2-8 z+4}{y^2 (z-1)^3}\,, \nonumber \\
& R_{10}=\frac{10 y^3-y^2 z^2+13 y^2 z-12 y^2-6 y z^2+12 y z-6 y-8 z^3+24 z^2-24 z+8}{y^2 (z-1)^2 (y+z-1)^2}\,, \nonumber\\
& R_{11}=\frac{2 y^2+3 y z^2-3 y z+2 y+3 z^3-9 z^2+10 z-4}{y (z-1)^2 (y+z-1) (y+z)}\,, \nonumber\\
& R_{12}=\frac{y^2 z+4 y z^2-6 y z+4 y+3 z^3-4 z^2+2 z}{y^2 (z-1) (y+z)^2}\,, \nonumber\\
& R_{13}=\frac{5 y^3 z+5 y^3+8 y^2 z^2+10 y^2 z+2 y^2+3 y z^3-5 y z^2+18 y z-6 y-8 z^3+12 z^2-4 z}{y (z-1)^2 z (y+z)^2} \,, \nonumber
 \end{align}
and
\begin{align}
 Q_1=&-12 \zeta (3) G(1-z,y)-\frac{2}{3} \pi ^2 G(1,z) G(1-z,y)-4 G(0,0,1,z) G(1-z,y) \nonumber \\ 
 &-4 G(1,0,1,z) G(1-z,y)+4 G(0,1,z) G(1-z,0,y)+\frac{2}{3} \pi ^2
   G(1-z,0,y) \nonumber \\ 
   &-4 G(0,1,z) G(1-z,1-z,y)-\frac{2}{3} \pi ^2 G(1-z,1-z,y)-4 G(1,1,z) G(-z,0,y) \nonumber \\ 
   &+4 G(0,1,z) G(-z,1-z,y)+4 G(0,1,z) G(-z,-z,y)+4 G(1,1,z) G(-z,-z,y) \nonumber \\ 
   &+4 G(0,1,1,z) G(-z,y)+4 G(1,0,1,z) G(-z,y)
   +4 G(1,z) \Big[ G(1-z,0,-z,y)\nonumber \\
   &- G(1-z,1-z,-z,y) - G(-z,0,1-z,y)- G(-z,1-z,0,y) \nonumber \\
   &+
   G(-z,1-z,-z,y) - G(-z,-z,0,y)+ G(-z,-z,1-z,y)+ G(-z,-z,-z,y) \Big]\nonumber \\
   &-4 G(1-z,0,1,0,y)+4 G(1-z,0,-z,1-z,y)+4 G(1-z,1-z,1,0,y)\nonumber \\
   &-4
   G(1-z,1-z,-z,1-z,y)+4 G(-z,0,1,0,y)-4 G(-z,0,1-z,1-z,y)\nonumber \\
   &-4 G(-z,1-z,0,1-z,y)+4 G(-z,1-z,1,0,y)-4 G(-z,1-z,1-z,0,y)\nonumber \\
   &+4 G(-z,1-z,-z,1-z,y)-4 G(-z,-z,0,1-z,y)-4
   G(-z,-z,1-z,0,y)\nonumber \\
   &+4 G(-z,-z,1-z,1-z,y)+4 G(-z,-z,-z,1-z,y)\,, \nonumber \\
 Q_2=&\frac{1}{2} G(1-z,y)+\frac{1}{2} G(1,z)-1\,, \nonumber\\
 Q_3=&\frac{1}{2} G(0,y) G(0,z)-\frac{1}{2} G(1,z) G(-z,y)-\frac{1}{2} G(-z,1-z,y)\nonumber \\
   &-\frac{1}{2} G(0,1,z)-\frac{1}{2} G(1,0,z)\,, \nonumber\\
 Q_4=&G(1,z) G(0,1-z,y)-G(1,z) G(0,-z,y)+G(1,z) G(1-z,0,y)\nonumber \\
   &+G(1,z) G(-z,0,y)-\frac{8}{3} G(1-z,y)-G(0,1,z) G(1-z,y)-G(0,1,z) G(-z,y)\nonumber \\
   &+G(0,z) G(0,1-z,y)-G(1,0,z)
   G(-z,y)+G(0,y) G(1,1,z)\nonumber \\
   &-G(0,z) G(1-z,1-z,y)-G(0,z)
   G(-z,1-z,y)+G(0,1-z,1-z,y)\nonumber \\
   &-G(0,-z,1-z,y)+G(1-z,0,1-z,y)-G(1-z,1,0,y)+G(1-z,1-z,0,y)\nonumber \\
   &+G(-z,0,1-z,y)+G(-z,1-z,0,y)+\frac{2}{3} G(1,0,y)-G(0,1,0,y)-\frac{8}{3}
   G(1,z)\nonumber \\
   &-\frac{2}{3} G(1,0,z)-G(0,1,1,z)-G(1,0,1,z)\,, \nonumber\\
 Q_5=&2 G(1-z,y)+2 G(1,z)\,,\nonumber \\
 Q_6=&-2 G(1,z) G(1-z,y)+\frac{4}{3} G(1-z,y)-2 G(1-z,1-z,y)+\frac{5}{3} G(1,0,y)\nonumber \\
   &+\frac{4}{3} G(1,z)+\frac{1}{3} G(1,0,z)-2 G(1,1,z)\,, \nonumber\\
 Q_7=&-\frac{1}{2} G(0,y) \,,\nonumber\\
 Q_8=&G(0,1,z) (-G(1-z,y))-G(0,1,z) G(-z,y)+G(1,z) G(0,1-z,y)\nonumber \\
   &+G(0,y) G(1,1,z)-G(1,1,z) G(-z,y)+G(1,z) G(1-z,0,y)\nonumber \\
   &-G(1,z) G(1-z,-z,y)+G(1,z) G(-z,0,y)-G(1,z)
   G(-z,1-z,y)\nonumber \\
   &-G(1,z)G(-z,-z,y)+G(0,1-z,1-z,y)+G(1-z,0,1-z,y)\nonumber \\
   &-G(1-z,1,0,y)+G(1-z,1-z,0,y)-G(1-z,-z,1-z,y)\nonumber \\
   &+G(-z,0,1-z,y)+G(-z,1-z,0,y)-G(-z,1-z,1-z,y)\nonumber \\
   &-G(-z,-z,1-z,y)-G(0,1,0,y)-G(0,1,1,z)-G(1,0,1,z)\,, \nonumber\\
 Q_9=&G(0,y) (-G(0,1,z))+\frac{1}{6} \pi ^2 G(1-z,y)+G(0,1,z) G(1-z,y)\nonumber \\
   &-G(1,z) G(0,-z,y)+G(1,z)G(1-z,-z,y)-G(0,-z,1-z,y)\nonumber \\
   &-G(1-z,1,0,y)+G(1-z,-z,1-z,y)-\frac{1}{6} \pi ^2 \Big[ G(0,y) - G(1,z)\Big]\nonumber \\
   &+G(0,1,0,y) +G(0,0,1,z)+G(1,0,1,z)+3 \zeta (3) \,,\nonumber\\
 Q_{10}=&-\frac{1}{2} G(1,z) G(-z,y)-\frac{1}{2} G(-z,1-z,y)+\frac{1}{2} G(1,0,y)-\frac{1}{2} G(0,1,z)-\frac{\pi ^2}{12}\,, \nonumber\\
 Q_{11}=&\frac{1}{2} G(0,z)\,, \nonumber\\
 Q_{12}=&-\frac{1}{2} G(0,z) G(1-z,y)+\frac{1}{2} G(1,z) G(-z,y)+\frac{1}{2} G(-z,1-z,y) \,,\nonumber\\
 Q_{13}=&\frac{1}{2} G(0,y) G(1,z)-\frac{1}{2} G(1,z) G(-z,y)+\frac{1}{2} G(0,1-z,y)+\frac{1}{2} G(1-z,0,y)\nonumber \\
   &-\frac{1}{2} G(-z,1-z,y)-\frac{1}{2}
   G(1,0,y)-\frac{1}{2} G(0,1,z)\,. \nonumber
\end{align}
\newline

\noindent For $K_2$ we write instead
$$K_2 = \sum_{i=14}^{19} R_i Q_i$$
with
\begin{align}
& R_{14}=\frac{y-z+2}{y (z-1) (y+z)} \,, \quad R_{15}=\frac{2 z-y}{(y+z)^4} \,, \quad R_{16}=\frac{2 y+2 z-3}{(y+z-1) (y+z)^2}\,, \nonumber \\
& R_{17}=\frac{(y-1) \left(y^3+y^2 z-7 y^2-y z^2-5 y z+5 y-z^3+2 z^2-z\right)}{y (y+z-1)^2 (y+z)^3}\,, \nonumber\\
& R_{18}=\frac{y^3+y^2 z-3 y^2-y z^2+3 y z+2 y-z^3+6 z^2-4 z}{(y+z-1)^2 (y+z)^3}\,, \nonumber\\
& R_{19}=\frac{y^3+4 y^2 z^2-5 y^2 z+3 y^2+4 y z^3-9 y z^2+10 y z-4 y+3 z^3-5 z^2+2 z}{y (z-1)^2 (y+z-1) (y+z)^2}\,, \nonumber
\end{align}
and
\begin{align}
Q_{14}=& 1 \,, \nonumber \\
 Q_{15}=&-\frac{2}{3} \pi ^2 G(1-z,y)-4 G(0,y) G(1,0,z)+4 G(1,0,z) G(1-z,y)
 \nonumber \\ 
 &-4 G(0,z) G(1-z,0,y)+4 G(1-z,1,0,y)+4 G(0,1,0,y)-\frac{2}{3} \pi ^2 G(1,z)
 \nonumber \\&+4
   G(0,1,0,z)+4 G(1,1,0,z)\,,  \nonumber \\
 Q_{16}=&2 G(0,y)\,, \nonumber \\
 Q_{17}=&2 G(1,0,z)-2 G(1,0,y)\,, \nonumber \\
 Q_{18}=&2 G(0,y) G(0,z)-4 G(1,0,z)+\frac{\pi ^2}{3}\,, \nonumber \\
 Q_{19}=&-G(0,z)  \,. \nonumber
\end{align}

As discussed above, the analytic results for the remaining coefficients can be found in the ancillary files.

\subsection{Helicity amplitudes for a Standard Model vector boson}
We are now ready to build the helicity amplitudes for a
standard model vector boson $V$ connecting a $q\bar{q}g$ system and to a lepton-antilepton pair.
We write  the coupling of the vector boson $V$ to two fermions $f_1 f_2$ 
in the two equivalent ways
\begin{align}
    -i e\, \Gamma_\mu^{V f_1 f_2} =& 
    - i \sqrt{4 \pi \alpha}\, \left[ v^V_{f_1 f_2} \gamma^\mu + a^V_{f_1 f_2} \gamma^\mu \gamma_5 \right] \nonumber \\
    =&-i \sqrt{4 \pi \alpha}\, \left[L_{f_1 f_2}^V \gamma^\mu \left( \frac{1-\gamma_5}{2} \right) + R_{f_1 f_2}^V \gamma^\mu \left( \frac{1+\gamma_5}{2} \right) \right]
    \label{eq:ewvertex}
\end{align}
where clearly
$$L_{f_1 f_2}^V = v^V_{f_1 f_2} - a^V_{f_1 f_2} \,, \quad
R_{f_1 f_2}^V = v^V_{f_1 f_2} + a^V_{f_1 f_2} $$
and for the three types of vector bosons we have
\begin{align}
    L_{f_1 f_2}^\gamma = R_{f_1 f_2}^\gamma = - e_{f_1} \delta_{f_1 f_2}
    \end{align}
\begin{align}
    L_{f_1 f_2}^Z = \frac{I_3^{f_1} - \sin^2{\theta_w} e_{f_1}}{\sin{\theta_w} \cos{\theta_w}} \delta_{f_1 f_2} \,, \qquad
    R_{f_1 f_2}^Z = -\frac{\sin{\theta_w} e_{f_1}}{\cos{\theta_w}} \delta_{f_1 f_2} 
\end{align}    
    \begin{align}
    L_{f_1 f_2}^W = \frac{\epsilon_{f_1,f_2}}{\sqrt{2}\sin{\theta_w}} \,, \qquad
    R_{f_1 f_2}^W = 0 \,.
\end{align}  
In the formulas above $\alpha$ is the electroweak coupling constant,
$\theta_w$ is the Weinberg angle, $I_3 = \pm 1/2$ is the third component
of the weak isospin and in all formulas the charges $e_i$
are measured in terms of the fundamental electric charge $e>0$.
Moreover, $\epsilon_{f_1,f_2}=1$ if $f_1 \neq f_2$ but belonging to the same
isospin doublet and zero otherwise. 
Finally, we write the propagator of the  vector boson $V$ of momentum $q$ and 
mass $m_V$ as $P_{\mu \nu}(q,m_V)$, whose expression in 
Lorentz gauge reads
\begin{equation}
    P_{\mu \nu}(q,m_V) = \frac{i \left( -g_{\mu \nu} + \frac{q_\mu q_\nu}{q^2}  \right)}{D\left(q^2,m_V^2\right)}\,.
\end{equation}
with
$$D\left(q^2,m_V^2\right) = q^2 - m_V^2 + i \Gamma_V m_V\,.$$

From (\ref{eq:ewvertex}), we can immediately read off 
the coupling factors of all non-singlet contributions: $v^V_{f_1 f_2}$ for the vector form factors and $a^V_{f_1 f_2}$ for the axial-vector form factors. Consequently, 
when re-casting the form factors into helicity 
amplitudes, we expect each amplitude for a specific 
quark helicity to contain a linear combination  
 of vector and axial-vector form 
factors. 

In the pure-singlet contributions, the vertex (\ref{eq:ewvertex}) is coupled to an internal quark loop, thus requiring a summation over the internal quark flavours. In the vector case, this summation 
amounts to an overall factor
\begin{align}
    N_{f,\mathcal{\gamma}}^{v} =\sum_q e_q \,, \qquad
     N_{f,Z}^{v} =\sum_q \frac{\left( L_{qq}^Z  + R_{qq}^Z\right)}{2}\,, \qquad
     N_{f,W}^{v} = 0\,, \label{eq:Nfv}
\end{align}
with the sums running over the active quark flavours. 
In the axial-vector case, the summation must be performed 
over complete quark isospin doublets. For mass-degenerate quarks in the doublet, the summation of up-type and down-type contribution cancels identically. In case of 
a mass-splitting in the doublet, the axial-vector pure-singlet contribution is obtained as the difference between up-type and down-type quark contributions in the loop, multiplied by a coupling factor 
\begin{align}
N_{f,\gamma}^{a} = 0\,, \qquad 
     N_{f,Z}^{a} =  \frac{1}{4 \sin{\theta_w} \cos{\theta_w}}  \,,
     \qquad
     N_{f,W}^{a} = 0\,. \label{eq:Nfa}
\end{align}
The last identity in~\eqref{eq:Nfv} 
is a consequence of charge conservation  while,
as already discussed, there can be a contribution from the 
axial-vector coupling 
to the pure-singlet amplitude only in the case
of the production of a $Z$ boson.
\newline

With these definitions, we write the helicity amplitudes for a vector 
boson $V$ as 
\begin{align}
	\mathcal{M}_{L+L}^V &=- 
 \frac{i \sqrt{4 \pi \alpha_s} (4 \pi \alpha)\, L_{l_5 l_6}^{V} }{\sqrt{2} \,D(p_{56}^2,m_V^2)}\, \mathbb{T}^{a}_{ij}
	 \bigg[ \langle 1 2 \rangle [1 3]^2 \Big( A_1 \langle 536] 
	+ A_2 \langle 526] \Big)
        + A_3 \langle 25 \rangle [13] [36] \bigg] \,,
	 \label{eq:HALmL} \\
	\mathcal{M}_{L-L}^V &=  - 
  \frac{i \sqrt{4 \pi \alpha_s} (4 \pi \alpha)\, L_{l_5 l_6}^{V}}{\sqrt{2}\, D(p_{56}^2,m_V^2)}\,  \mathbb{T}^{a}_{ij}
  \bigg[ 
  \langle 23 \rangle^2 [1 2] \Big( B_1 \langle 536] 
	+ B_2 \langle 516] \Big)
        + B_3 \langle 23 \rangle \langle 35 \rangle [16]  
\bigg]\,, \label{eq:HALpL}
 \end{align}
where $p_{56} = p_5 + p_6$.
The $l$-loop coefficients can be written in terms of the $l$-loop 
$\Omega_i$ and $\Lambda_i$ 
introduced above\footnote{See discussion around~\eqref{eq:OmLam}.} as follows 
\begin{align}
A_i^{(l)} &=  L_{q_1 q_2}^V \alpha_i^{(l),n} 
+ N_{f,V}^v \alpha_i^{(l),p} + N_{f,V}^a \left( \beta_i^{(l),p0} - \beta_i^{(l),pm} \right)\,, \\
B_i^{(l)} &=  L_{q_1 q_2}^V \gamma_i^{(l),n} 
+ N_{f,V}^v \gamma_i^{(l),p} + N_{f,V}^a \left( \delta_i^{(l),p0} - \delta_i^{(l),pm} \right) \,,
\end{align}
where we used  the fact that the non-singlet vector and axial-vector
parts are equal after UV renormalisation and IR subtraction.

In the same way, be obtain for right-handed quarks

\begin{align}
	\mathcal{M}_{R+L}^V &=
 \frac{i \sqrt{4 \pi \alpha_s} (4 \pi \alpha)\, L_{l_5 l_6}^{V} }{\sqrt{2} \,D(p_{56}^2,m_V^2)}\,  \mathbb{T}^{a}_{ij}
	  \bigg[ [23]^2 \langle 1 2\rangle \Big( C_1 \langle 536] 
	+ C_2 \langle 516] \Big)
        + C_3 [23] [36] \langle 15 \rangle \bigg] \,,
	 \label{eq:HARmL} \\
	\mathcal{M}_{R-L}^V &=  
  \frac{i \sqrt{4 \pi \alpha_s} (4 \pi \alpha)\, L_{l_5 l_6}^{V}}{\sqrt{2}\, D(p_{56}^2,m_V^2)}\,  \mathbb{T}^{a}_{ij}
  \bigg[ [1 2] \langle 1 3\rangle^2 \Big( D_1 \langle 536] 
	+ D_2 \langle 526] \Big)
        + D_3 [26] \langle 13\rangle \langle 35\rangle \bigg]\,, \label{eq:HARpL}
 \end{align}
with
\begin{align}
C_i^{(l)} &=  R_{q_1 q_2}^V \gamma_i^{(l),n} 
+ N_{f,V}^v \gamma_i^{(l),p} - N_{f,V}^a \left( \delta_i^{(l),p0} - \delta_i^{(l),pm} \right) \,, \\
D_i^{(l)} &=  R_{q_1 q_2}^V \alpha_i^{(l),n} 
+ N_{f,V}^v \alpha_i^{(l),p} - N_{f,V}^a \left( \beta_i^{(l),p0} - \beta_i^{(l),pm} \right)\,.
\end{align}

We stress that the remaining four helicity amplitudes can be obtained from the ones given above, by a CP tranformation \emph{and changing all relevant couplings}. The CP transformation acts on the spinor products by swapping angle and square brackets, but it leaves the functional part of the coefficients unchanged.

\subsection{Checks on the results}
In deriving the 
results for the renormalized form factors in Section~\ref{sec:renorm}
and the resulting helicity amplitudes in this section, 
we have performed various checks, which we briefly 
summarize in the following. 

The infrared pole structure of the form factors 
at one loop and two loops can be predicted in terms 
of universal IR pole operators and lower-order results, 
as described in~(\ref{eq:nsfinite},\ref{eq:psfinite}). We observe 
that all form factors up to two loops 
reproduce the predicted IR pole structure. 
This is a particularly strong check for what concerns the axial-vector part of the result, since the latter requires non-trivial UV renormalisation, which depends on the scheme used to deal with $\gamma_5$ in dimensional regularisation.

As a second non-trivial check for our correct implementation of the Larin scheme, we have verified explicitly up to two loops that the non-singlet axial-vector and purely vector helicity amplitudes agree after UV renormalisation and IR subtraction,  see~\eqref{eq:NSavv}.

The vector parts of the helicity amplitudes
were computed previously up to two loops~\cite{Garland:2002ak}, and we reproduced these 
earlier results. 
The axial-vector singlet parts were only known to 
one-loop, and  
 we verified that the 
 resulting  one-loop helicity amplitudes agree with the literature~\cite{Bern:1997sc}. 
 
 Finally, we performed a thorough check of the helicity amplitudes up to one loop against \OLL~\cite{Cascioli:2011va,Buccioni:2019sur}, 
 both for the exchange of a virtual photon and of a $Z$ boson. This allowed us 
 to validate all electroweak couplings, including the 
 overall normalisation of $N_{f,Z}^a$, see \eqref{eq:Nfa}.

\section{Conclusions}
\label{sec:conc}

The two-loop helicity amplitudes for $V\to q\bar q g$ 
constitute the purely virtual contribution to the NNLO corrections to $e^+e^- \to 3$~jets, as well as 
in different kinematical crossings 
to $ep \to (2+1)$~jets and $pp\to V+1$~jet. They were computed already long ago~\cite{Garland:2001tf,Garland:2002ak} for a purely vector-like coupling of the boson 
$V$. Owing to chirality conservation for massless fermions, these results can be extended in a straightforward manner to an axial-vector coupling of $V$~\cite{Garland:2002ak,Gehrmann:2011ab} for non-singlet type contributions, where the vector boson couples to the external quark line. 
These amplitudes were used  subsequently in 
the NNLO calculations for the above processes in  $e^+e^-$ annihilation~\cite{Gehrmann-DeRidder:2007vsv,Weinzierl:2009ms,DelDuca:2016csb}, deeply inelastic electron-proton collisions~\cite{Currie:2017tpe} and proton-proton collisions~\cite{Gehrmann-DeRidder:2015wbt,Boughezal:2015ded,Neumann:2022lft}. 
\hyphenation{re-derive}

In this paper, we complete the computation of the 
two-loop 
$V\to q\bar q g$ helicity amplitudes by deriving the 
previously missing pure-singlet axial-vector contributions. Our calculation is enabled by the construction of a new four-dimensional tensor basis~\cite{Peraro:2019cjj,Peraro:2020sfm} for the $V\to q\bar q g$
amplitudes, which avoids the introduction of evanescent tensor structures and allows a consistent application of the Larin scheme for axial couplings in dimensional regularization in the computation of the associated form factors. In this new basis, we rederive the non-singlet helicity amplitudes at two loops, confirming the earlier results~\cite{Garland:2001tf,Garland:2002ak} and explicitly demonstrating the equivalence of vector and axial-vector amplitudes in the non-singlet case, after renormalization and IR factorization. 

In the pure-singlet axial-vector amplitudes, the vector boson couples to an internal quark loop. These amplitudes are affected by the axial anomaly, which cancels in the electroweak Standard Model upon summation over weak isospin doublets. We compute the two-loop pure-singlet axial-vector form factors, separately for massless internal quarks and in a large-mass expansion for massive internal quarks, and recovering in both cases the correct universal divergent behaviour of the axial anomaly. Subleading terms in the large-mass expansion are also computed to demonstrate the internal consistency of the approach. The combination of massless and massive internal quarks describes the pure-singlet axial-vector contribution from top-bottom mass splitting. 

By combining our newly derived two-loop pure-singlet 
axial-vector amplitudes with the other pure-singlet 
contributions that contribute at the same order~\cite{vanderBij:1988ac,Hopker:1993pb,Bern:1997sc}, 
it will now be possible to consistently compute the 
axial-vector pure-singlet contributions to $V+$jet production at NNLO. Moreover, in combination with the 
recently derived three-loop pure-singlet axial-vector quark form factors~\cite{Gehrmann:2021ahy,Chen:2021rft}, these results can be extended to differential cross sections in $Z$ boson production at N3LO.  

Besides their potential relevance for differential lepton pair distributions in $Z$ and $Z$+jet production processes at hadron colliders, the pure-singlet axial-vector contribution could also have an impact on three-jet production observables at $e^+e^-$ colliders, especially on the event orientation~\cite{Gehrmann:2017xfb} and on oriented event shapes derived from it.

\section*{Acknowledgments}
We are grateful to Federico Buccioni for his assistance and patience in helping us check our results against \OLL, and to Fabrizio Caola for useful comments on the manuscript.
This work was supported in part by  the Excellence
Cluster ORIGINS funded by the Deutsche Forschungsgemeinschaft 
(DFG, German Research Foundation) under Germany's Excellence Strategy -- EXC-2094-390783311,
by the Swiss National Science Foundation (SNF) under contract 200020-204200, and by the European Research Council (ERC) under the European Union's
 research and innovation programme grant agreements 949279 (ERC Starting Grant HighPHun), 101040760 (ERC Starting Grant FFHiggsTop) and 
 101019620 (ERC Advanced Grant TOPUP).
\appendix

\bibliography{Biblio}

\end{document}